\documentclass[pra, floatfix, twocolumn]{revtex4-2}
\usepackage{graphicx}
\usepackage{xcolor}
\usepackage{float}
\usepackage{amsmath, amsfonts, amssymb, bm, dsfont, mathtools}
\usepackage[bb=libus]{mathalpha}
\usepackage{comment}
\usepackage{hyperref}
\hypersetup{
	colorlinks=true,
	linkcolor=blue,
	citecolor=red,
	filecolor=magenta,
	urlcolor=purple,
}

\usepackage[switch]{lineno} % enable line number

\usepackage[normalem]{ulem}
\begin{document}
	%\linenumbers % enable line number
%%%%%%%%%%%%%%%%%%%%%%%%%%%%%%%%%%%%%%%%%%%%%%%%%%%%%%%%%%%%%%%%%
\title{Effects of quadratic frequency chirping and carrier-envelope phase on \\ Schwinger pair production in polarized electric fields}
\author{Abhinav Jangir}
\email{2022rpy9087@mnit.ac.in}
\address{Department of Physics, MNIT Jaipur, Jaipur, Rajasthan, India}
\date{\today}
%%%%%%%%%%%%%%%%%%%%%%%%%%%%%%%%%%%%%%%%%%%%%%%%%%%%%%%%%%%%%%%%%%%
\begin{abstract}
	We investigate electron-positron pair production in spatially homogeneous, time-dependent polarized electric fields using the Dirac-Heisenberg-Wigner formalism, focusing on the effects of symmetric and asymmetric quadratic frequency chirping and the carrier-envelope phase (CEP). Our analysis reveals how these field parameters govern both the momentum-space structure and the overall yield of the produced pairs. Frequency chirping substantially reshapes the momentum distributions by modifying their interference patterns and peak occupations. The momentum distributions are particularly sensitive to weak chirping, whereas strong chirping produces pronounced oscillatory and ring-like structures across all polarization states. The total pair yield increases with chirp strength, with symmetric chirping consistently producing a higher yield than asymmetric chirping for the parameter range considered. Moreover, the sensitivity of the total number density to polarization decreases progressively with increasing chirp strength, becoming negligible in the strongly chirped regime. The CEP further modifies the momentum distributions by altering their shapes and relative peak occupations, with the extent of these changes depending on the polarization state. Our findings establish a theoretical framework for exploring external-field configurations that optimize the yield of Schwinger pair production.
\end{abstract}
%%%%%%%%%%%%%%%%%%%%%%%%%%%%%%%%%%%%%%%%%%%%%%%%%%%%%%%%%%%%%%%%%%%
\maketitle
%%%%%%%%%%%%%%%%%%%%%%%%%%%%%%%%%%%%%%%%%%%%%%%%%%%%%%%%%%%%%%%%%
\section{Introduction}\label{sec:Intro}
%%%%%%%%%%%%%%%%%%%%%%%%%%%%%%%%%%%%%%%%%%%%%%%%%%%%%%%%%%%%%%%%%
	Vacuum electron-positron ($e^+e^-$) pair production is a fundamental nonperturbative prediction of quantum electrodynamics (QED), describing the conversion of vacuum fluctuations into real particles under sufficiently strong electromagnetic fields. The theoretical foundations of this phenomenon emerged from Dirac's relativistic wave equation, which led to the prediction of the positron and established the basis for a relativistic description of particle–antiparticle states \cite{dirac_quantum_1928}. Subsequent developments by Sauter provided an early quantum-mechanical treatment of field-induced barrier penetration \cite{sauter_behavior_1931}, while Heisenberg and Euler formulated the effective action describing the nonlinear response of the quantum vacuum to electromagnetic fields \cite{heisenberg_folgerungen_1936}. In 1951, Schwinger employed the proper-time formalism to derive the nonperturbative pair-production rate in a constant electric field, establishing vacuum breakdown as a fundamental consequence of QED \cite{schwinger_gauge_1951}. The corresponding phenomenon, commonly known as the Schwinger effect, becomes significant when the electric-field strength approaches the critical value $E_{\mathrm{cr}}=m^2c^3/(e\hbar)\approx 1.3\times10^{18}\,\mathrm{V/m}$ ($m$ and $-e$ are the mass and charge of the electron, respectively), at which the work performed by the field over a Compton wavelength becomes comparable to the electron rest energy \cite{fradkin_quantum_1991}.

	Despite its fundamental importance, direct experimental observation of the Schwinger effect remains elusive because the critical electric-field strength exceeds the capabilities of currently available high-intensity laser systems. Nevertheless, advances in laser technology and ongoing efforts at major international facilities have intensified research into strong-field QED phenomena \cite{eli, xfel, yanovsky_ultra-high_2008, pike_photonphoton_2014, marklund_quantum_2009}. Several studies have suggested that the intensities achievable with next-generation ultrastrong laser facilities could make the experimental observation of vacuum pair production possible \cite{blaschke_pair_2006, bell_pissibility_2008, alkofer_pair_2001, ringwald_pair_2001}. Particular attention has been devoted to mechanisms that can enhance vacuum pair production at subcritical field strengths, where the production rate is exponentially suppressed and direct observation remains challenging \cite{bulanov_multiple_2010, di_piazza_barrier_2009, dunne_catalysis_2009, schutzhold_dynamically_2008, titov_enhanced_2012}.
	
	One such mechanism is the dynamically assisted Schwinger effect (DASE), in which a slowly varying, strong electric field is combined with a weaker, rapidly varying field to enhance vacuum pair production \cite{schutzhold_dynamically_2008, schneider_dynamically_2016, linder_pulse_2015}. The substantial enhancement of pair production offered by this mechanism has motivated extensive investigations across a broad range of field parameters and configurations, encompassing both temporally and spatially inhomogeneous backgrounds \cite{nuriman_enhanced_2012, li_study_2021, jangir_dynamically_2026, li_enhanced_2021, ababekri_effects_2019, blaschke_lifting_2015, fey_momentum_2012, orthaber_momentum_2011, bras_relative_2026, aleksandrov_dynamically_2018, linder_pulse_2015}.

	Researchers have developed a wide range of analytical and numerical methods to describe nonperturbative quantum dynamics in external electromagnetic fields. Semiclassical techniques, particularly the worldline-instanton method \cite{esposti_worldline_2023, esposti_momentum_2024, gies_pair_2005, dunne_worldline_2005, dunne_worldline_2006, schneider_dynamically_2016, dumlu_complex_2011}. Complementary descriptions are available through effective-Lagrangian approaches \cite{dunne_heisenberg-euler_2005} and the Wentzel-Kramers-Brillouin approach \cite{brezin_pair_1970, popov_pair_1972, oertel_wkb_2019}. Among the computational frameworks, the Furry-picture formulation \cite{aleksandrov_momentum_2017, aleksandrov_pair_2020} as well as the quantum kinetic theory including formulations based on the quantum Vlasov equation \cite{kluger_quantum_1998, alkofer_pair_2001, nuriman_enhanced_2012, abdukerim_effects_2013, dumlu_interference_2011, gong_electron-positron_2020, jangir_carrier-envelope_2026, dumlu_schwinger_2010}. Other approaches include computational quantum field theory \cite{braun_numerical_1999, li_study_2021}, the Dirac–Heisenberg–Wigner (DHW) formalism \cite{blinne_pair_2014, li_nonperturbative_2015, li_effects_2015, jangir_dynamically_2026} and the scattered-wave-function approach \cite{torgrimsson_momentum_2025, torgrimsson_solving_2025}. Collectively, these approaches provide distinct perspectives on vacuum decay, quantum tunneling, multiphoton processes, and the influence of field geometry on particle creation. For comprehensive discussions of strong-field QED, intense-laser interactions, and related developments in nuclear and particle physics, we refer readers to Refs. \cite{fedotov_advances_2023, di_piazza_extremely_2012}.
	
	The pair-production process is highly sensitive to external field parameters, including the field frequency, pulse duration, CEP, number of pulses, and spatial scale. A wide range of temporal and spatial field configurations has been investigated over various parameter regimes to identify conditions that optimize vacuum pair production \cite{nuriman_enhanced_2012, abdukerim_effects_2013, dumlu_interference_2011, dumlu_schwinger_2010, li_effects_2015, dumlu_stokes_2010, hebenstreit_momentum_2009, hebenstreit_optimization_2014, jangir_carrier-envelope_2026, jangir_dynamically_2026, olugh_asymmetric_2020, olugh_pair_2019}. Among the various approaches, frequency chirping has emerged as a promising means of controlling pair-production dynamics. By introducing a time-dependent modulation of the instantaneous frequency, chirping can substantially alter both the total pair yield and the momentum distributions of the produced particles. Its effects have been investigated in spatially homogeneous, time-dependent electric fields \cite{gong_electron-positron_2020, dumlu_schwinger_2010, chen_asymmetric_2024}, spatially inhomogeneous oscillating backgrounds \cite{ababekri_chirp_2020, li_enhanced_2021, jangir_effects_2026, bake_vacuum_2025,mohamedsedik_phase_2023}, and polarized electric fields \cite{olugh_frequency_2026, olugh_pair_2019, jangir_dynamically_2026}. Beyond linear frequency modulation, recent studies have explored the effects of symmetric and asymmetric quadratic frequency chirping on pair production in spatially inhomogeneous electric fields \cite{chen_electron_2026, osman_efficient_2023}. The CEP is another important parameter governing pair-production dynamics. Its influence on the momentum distributions and total number density of produced particles has been extensively investigated in various studies \cite{bras_two_2020, banerjee_electron_2018, jansen_strong_2016, abdukerim_effects_2013, jangir_carrier-envelope_2026, mohamedsedik_phase_2023, krajewska_phase_2012, dumlu_schwinger_2010}.
	
	Despite extensive investigations of frequency chirping and CEP effects in vacuum pair production, their interplay in spatially homogeneous, time-dependent polarized electric fields remains insufficiently explored, particularly in the context of quadratic frequency modulation. In this work, we investigate vacuum $e^+e^-$ pair production in polarized electric fields, considering both symmetric and asymmetric quadratic frequency chirping. Within the DHW formalism, we systematically examine how the chirping configuration and field polarization jointly influence the total pair yield and momentum distributions of the produced particles. We further investigate the effects of the CEP on the momentum-space structure of the produced pairs for different polarization states, elucidating how the phase of the carrier wave modifies the momentum distributions in rotating and elliptically polarized fields. By addressing these aspects within a unified framework, this study provides a comprehensive analysis of the interplay between quadratic frequency chirping, polarization, and CEP in vacuum pair production.
	
	The remainder of this paper is organized as follows. In Sec. \ref{sec:Theory}, we introduce the electric-field model and outline the DHW formalism used in our numerical calculations. Section \ref{sec:polarization and chirp effects} presents the numerical results, focusing on the effects of frequency chirping for different field polarizations. In Sec. \ref{sec:CEP effects}, we investigate the influence of the carrier-envelope phase on pair production. Finally, Sec. \ref{sec:Conclusion} summarizes our principal findings and presents our conclusions.
%%%%%%%%%%%%%%%%%%%%%%%%%%%%%%%%%%%%%%%%%%%%%%%%%%%%%%%%%%%%%%%%%
\section{Background field and Theoretical formalism}\label{sec:Theory}
%%%%%%%%%%%%%%%%%%%%%%%%%%%%%%%%%%%%%%%%%%%%%%%%%%%%%%%%%%%%%%%%%
	%--------------------------------------------------------------------------
	\subsection{Model of the background electric field}\label{subsec:field_model}
\begin{figure}[tbh]
		\centering
		\includegraphics[width=\columnwidth]{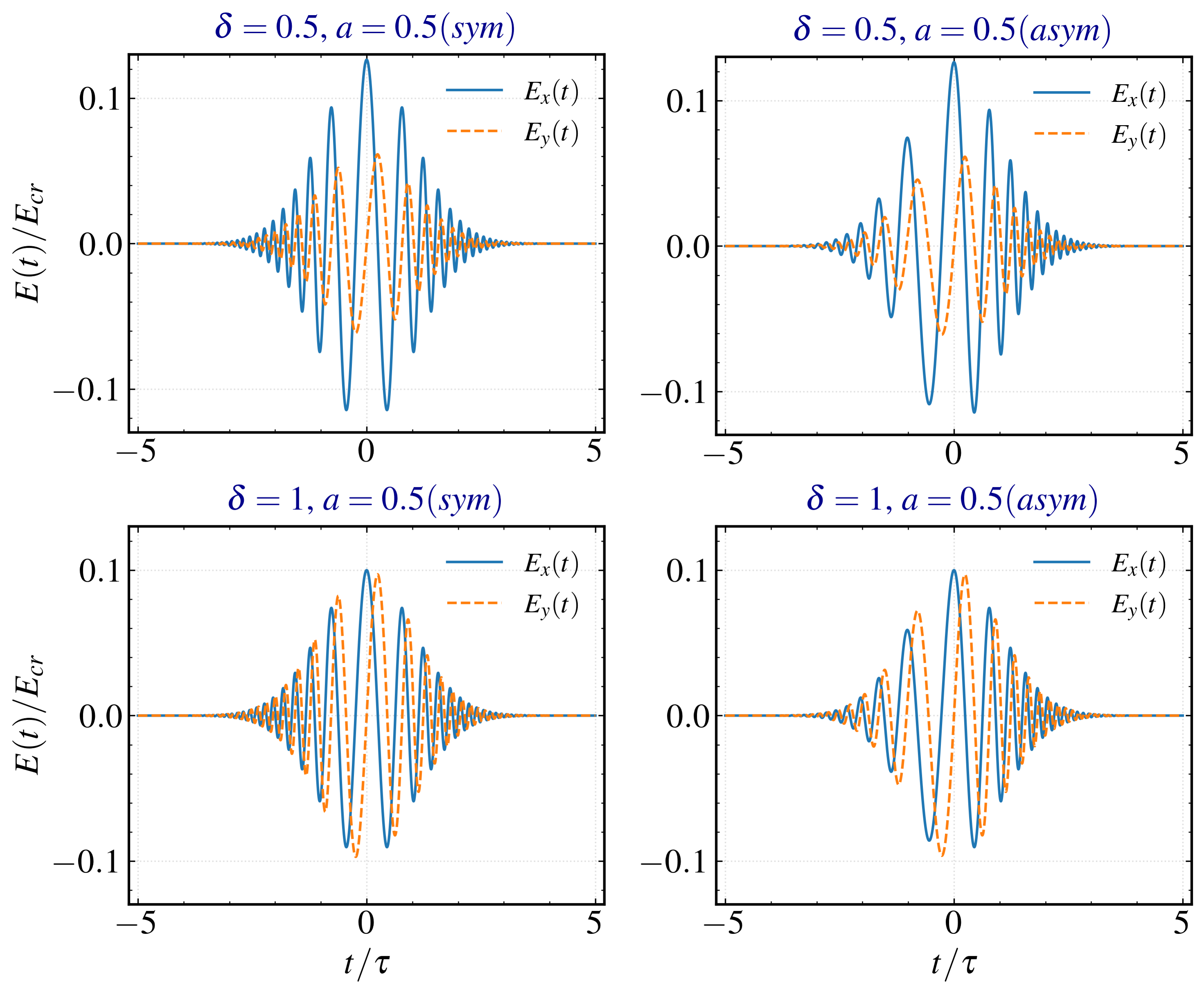}
		\caption{Electric-field configuration defined by Eq.~\eqref{eq:field_model}. The panels show the temporal evolution of the field components $E_x(t)$ and $E_y(t)$ for different polarization states, considering both \textit{symmetric} and \textit{asymmetric} chirp configurations at fixed CEP $\varphi=0$. The chirp parameters are $b_1=a\omega/(2\tau)$ and $b_2=a\omega/(2\tau^2)$, with the dimensionless chirp-control parameter fixed at $a=0.5$. The remaining field parameters are fixed to the values given in Eq.~\eqref{eq:field_parameters}.}
		\label{fig:field}
	\end{figure}

	%--------------------------------------------------------------------------
	To investigate the influence of frequency chirping and polarization on $e^+e^-$ pair production, we consider a spatially homogeneous, time-dependent electric field in the $(x,y)$ plane. The field is modeled as a Gaussian pulse with a time-dependent phase, allowing independent control of polarization and frequency chirping. The electric field is given by
	\begin{equation}\label{eq:field_model}
			\mathbf{E}(t) = \dfrac{E_{0}}{\sqrt{1 + \delta^2}}\,\text{exp}\left( -\frac{t^2}{2\tau^2} \right)\begin{pmatrix}
				\cos[\Phi(t)] \\
				\delta \sin[\Phi(t)]\\
				0
			\end{pmatrix}.	
	\end{equation}
	Here, $E_0$ denotes the field strength and $\tau$ the pulse duration. The parameter $\delta$ controls the polarization. For $\delta=0$, the field is linearly polarized, while $\delta=\pm1$ corresponds to circular polarization. Intermediate values, $0<|\delta|<1$, describe elliptically polarized fields, with the sign of $\delta$ determining the sense of rotation of the electric-field vector. The factor $1/\sqrt{1+\delta^2}$ is introduced to maintain a consistent field-strength scale when comparing different polarization states. The phase $\Phi(t)$ contains the carrier frequency and the frequency-chirp contributions. We consider the following phase functions corresponding to symmetric and asymmetric quadratic chirping:
	\begin{align}
		\Phi_\mathrm{sym}(t) &= \omega t + b_1|t|t + b_2|t|^2t + \varphi,\\
		\Phi_\mathrm{asym}(t) &= \omega t + b_1t^2 + b_2t^3 + \varphi,
	\end{align}
	where $\omega$ is the original carrier frequency, while $b_1$ and $b_2$ characterize the chirp strength and are defined as $b_1 = a\omega/(2\tau)$ and $b_2 = a\omega/(2\tau^2)$, with $0 \le a \le 1$. The chirping is considered in the interaction region $-\tau \le t \le \tau$, over which $e^+e^-$ pair creation is most significant. The parameter $\varphi$ denotes the CEP. Figure \ref{fig:field} shows the time-evolution of the field profiles for different polarization states and chirping configurations.
	
	Throughout this paper, we employ natural units ($\hbar = c = 1$). Furthermore, in our numerical calculations, a set of parameters characterizing the electric field  \eqref{eq:field_model} are fixed as 
	\begin{equation}\label{eq:field_parameters}
		E_{0} = 0.1\sqrt{2}\,E_\text{cr},\quad \omega = 0.5\,m,\quad  \tau = 12/m.
	\end{equation}
	Pair production in a strong time-dependent electric field can proceed through two limiting mechanisms: nonperturbative tunneling, commonly referred to as the Schwinger mechanism, and perturbative multiphoton absorption. These regimes are characterized by the Keldysh parameter $\gamma = {m\omega}/{eE}$ \cite{keldysh_ionization_1965, brezin_pair_1970}. For $\gamma\ll1$, the field varies slowly compared with the characteristic tunneling time, and pair production occurs predominantly through the nonperturbative tunneling mechanism. In contrast, $\gamma\gg1$ corresponds to the multiphoton regime, where pair production proceeds predominantly through the absorption of multiple photons. The intermediate region, $\gamma\sim1$, represents a crossover between the two mechanisms. With the field parameters in Eq. \eqref{eq:field_parameters}, the unchirped field has $\gamma=3.54\sqrt{1 + \delta^2}$, placing it in the intermediate regime for all considered values of $\delta$. For chirped fields, the Keldysh parameter is evaluated using the effective frequency $\omega_{\mathrm{eff}}(t)=\omega+b_1t+b_2t^2$ for asymmetric chirping and $\omega_{\mathrm{eff}}(t)=\omega+b_1|t|+b_2|t|^2$ for symmetric chirping, with an upper bound of $2\omega$.

	The electric field defined in Eq.~\eqref{eq:field_model} represents an idealized, spatially homogeneous configuration adopted to investigate pair-production dynamics under controlled field conditions. Such a model can be regarded as the \textit{dipole approximation} to a realistic standing-wave configuration formed by two counterpropagating laser pulses, provided that the spatial variation of the field occurs over characteristic length scales much larger than the pair-formation length. Under this condition, spatial variations of the electromagnetic field across the pair-formation region can be neglected, allowing the pair-production dynamics to be studied by considering only the temporal evolution of the electric field. The applicability and limitations of this approximation in strong-field pair production have been discussed in Refs.~\cite{aleksandrov_momentum_2017, lv_role_2018}. Accordingly, the present study focuses on the spatially homogeneous limit and does not account for spatial gradients or other effects associated with the finite spatial extent of realistic laser fields. Extending the analysis to fully space- and time-dependent electromagnetic field configurations would provide a more complete description of pair production under realistic laser conditions and is left for future work.
	%--------------------------------------------------------------------------
	\subsection{Theoretical description : The DHW formalism} \label{subsec:DHW_formalism}
	%--------------------------------------------------------------------------
	We employ the real-time Dirac–Heisenberg–Wigner formalism, widely used to study vacuum pair production in strong external fields \cite{birula_phase_1991, hebenstreit_schwinger_2010, hebenstreit_particle_2011}. It provides phase-space information on pair-production dynamics for general background fields. Following Refs. \cite{kohlfurst_electron-positron_2015, hebenstreit_schwinger_2011, birula_phase_1991}, we outline the essential steps required for our numerical calculations, beginning with the gauge-covariant density operator of the Dirac field $\Psi$ in the Heisenberg picture.
	\begin{equation}\label{eq:density operator}
		\hat{\mathcal{C}}_{\alpha\beta}(r,s) = \mathcal{U}(A,r,s)[\bar{\Psi}_\beta(r-s/2), {\Psi}_\alpha(r+s/2)],
	\end{equation}
	where $r$ and $s$ are the center of mass and relative coordinates, respectively. The Wilson line factor
	\begin{equation}
		\mathcal{U}(A; r, s) = \exp \left[ ies \int_{-1/2}^{1/2} d\xi\, A(r + \xi s) \right].
	\end{equation}
	ensures the gauge invariance of the density operator by incorporating the background gauge field $A$ and the elementary charge $e$. The covariant Wigner operator is then obtained by Fourier transforming the density operator in Eq. \eqref{eq:density operator} with respect to the relative coordinate $s$
	\begin{equation}\label{eq:covariant Wigner operator}
		\hat{\mathcal{W}}_{\alpha\beta}(r, p) = \frac{1}{2} \int d^4s\, e^{ip s} \hat{\mathcal{C}}_{\alpha\beta}(r, s).
	\end{equation}
	For vacuum pair production, taking the vacuum expectation value of the covariant Wigner operator in Eq. \eqref{eq:covariant Wigner operator} yields the covariant Wigner function (or simply the Wigner function) $\mathbb{W}(r, p)$
	\begin{equation}\label{eq:covariant Wigner function}
		\mathbb{W}(r, p) = \langle \Phi | \hat{\mathcal{W}}(r, p) | \Phi \rangle.
	\end{equation}
	The Wigner function can be decomposed into Dirac bilinears, yielding 16 covariant Wigner coefficients.
	\begin{equation}\label{eq:Wigner function components}
		\mathbb{W} = \frac{1}{4} \left( \mathds{1} \mathbb{S} + i\gamma_5 \mathbb{P} + \gamma^\mu \mathbb{V}_\mu + \gamma^\mu \gamma_5 	\mathbb{A}_\mu + \sigma_{\mu\nu} \mathbb{T}^{\mu\nu} \right).
	\end{equation}
	Under orthochronous Lorentz transformations, the 16 Wigner coefficients transform as a scalar \(\mathbb{S}\), pseudoscalar \(\mathbb{P}\), vector \(\mathbb{V}_\mu\), axial vector \(\mathbb{A}_\mu\), and antisymmetric tensor \(\mathbb{T}_{\mu\nu}\), corresponding to mass density, pseudoscalar condensate, fermion current, polarization, and electric dipole-moment density, respectively. The Wigner function satisfies the equations of motion \cite{hebenstreit_schwinger_2011, kohlfurst_electron-positron_2015}:
	\begin{equation}\label{eq:dynamical EOM of W}
		D_t \mathbb{W} = -\frac{1}{2} \textbf{D}_x [\gamma^0 \bm{\gamma}, \mathbb{W}] + i m [\gamma^0, \mathbb{W}] - i \textbf{P} \{ \gamma^0 	\bm{\gamma}, \mathbb{W} \},
	\end{equation}
	where $D_t$, $\textbf{D}_x$, and $\textbf{P}$ denote the pseudodifferential operators.
	
	Substituting Eq.~\eqref{eq:Wigner function components} into Eq.~\eqref{eq:dynamical EOM of W} yields a set of partial differential equations for the 16 Wigner components. For the spatially homogeneous, time-dependent field in Eq.~\eqref{eq:field_model}, applying the method of characteristics \cite{blinne_pair_2014, blinne_comparison_2016} and replacing the kinetic momentum $\mathbf{p}$ with the canonical momentum $\mathbf{k}$ reduces the system to $10$ ordinary differential equations for the non-vanishing equal-time Wigner components, given by
	\begin{equation}
		\mathbb{w} = (\mathbb{s},\mathbb{v}_i,\mathbb{a}_i,\mathbb{t}_i),\quad \mathbb{t}_i:=\mathbb{t}_{0i} -\mathbb{t}_{i0}.
	\end{equation}
	Correspondingly, the non-vanishing vacuum initial conditions are 
	\begin{equation}
		\mathbb{s}_{\mathrm{vac}} = -\frac{2m}{\sqrt{\textbf{p}^2 + m^2}}, \qquad
		\mathbb{v}_{i,\mathrm{vac}} = -\frac{2p_i}{\sqrt{\textbf{p}^2 + m^2}}.	
	\end{equation}
	For the explicit forms of these 10 equations of motion, we refer readers to Refs. \cite{hebenstreit_schwinger_2011, kohlfurst_electron-positron_2015}. 
	
	The single-particle momentum distribution function is defined as
	\begin{equation}\label{eq:distribution function}
		f_\textbf{k}(t) = \frac{1}{2\Omega_\textbf{k}(t)} \left( \varepsilon - \varepsilon_{\mathrm{vac}} \right),
	\end{equation}
	where $\Omega_\mathbf{k}(t) = \sqrt{\mathbf{p}^2 + m^2} = \sqrt{(\mathbf{k} - e\mathbf{A})^2 + m^2}$, is the total energy of the particles, where $\mathbf{A}(t)$ denotes the vector potential corresponding to the electric field $\mathbf{E}(t)$ of Eq. \eqref{eq:field_model}. $\varepsilon = m \mathbb{s} + p_i \mathbb{v}_i$ denotes the phase space energy density. Further, we define an auxiliary three dimensional vector 
	\begin{equation}
		\textbf{v}_\textbf{k}(t) \coloneqq \mathbb{v}_{i\textbf{k}}(t) - (1 - f_\textbf{k}(t))\mathbb{v}_{i\textbf{k},\text{vac}}(t). 
	\end{equation}
	The one-particle momentum distribution function $f_\textbf{k}(t)$ is then obtained by solving the following ten ordinary differential equations (ODEs) for $f$ and the other nine auxiliary quantities $\textbf{v},\textbf{a}\coloneqq\mathbb{a}_i,\textbf{t}\coloneqq \mathbb{t}_i$:
	\begin{equation}\label{eq:DHW-10 ODEs}
		\begin{split}
			\dot{f} &= \frac{e\,\mathbf{E}\cdot\mathbf{v}}{2\,\Omega}, \\[6pt]
			\dot{\mathbf{v}} &= \frac{2}{\Omega^{3}}
			\Big[ (e\mathbf{E}\cdot\mathbf{p})\,\mathbf{p} - e\,\mathbf{E}\,\Omega^{2} \Big]\,(f-1)
			- \frac{(e\mathbf{E}\cdot\mathbf{v})\,\mathbf{p}}{\Omega^{2}}\\
			&\hspace{0.5cm} - 2\,\mathbf{p}\times\mathbf{a} - 2m\,\mathbf{t}, \\[6pt]
			\dot{\mathbf{a}} &= -\,2\,\mathbf{p}\times\mathbf{v}, \\[6pt]
			\dot{\mathbf{t}} &= \frac{2}{m}\Big[ m^{2}\mathbf{v} + (\mathbf{p}\cdot\mathbf{v})\,\mathbf{p} \Big],
		\end{split}
	\end{equation}
	with the initial conditions $f_\mathbf{k}(-\infty) = \mathbf{v}_\mathbf{k}(-\infty) = \mathbf{a}_\mathbf{k}(-\infty) = \mathbf{t}_\mathbf{k}(-\infty) = 0$, where an overdot represents the total time derivative. 
	
	The asymptotic momentum distribution of produced pairs is
	\begin{equation}
		f_\mathbf{k}(\infty) = \lim_{t\rightarrow+\infty}f_\mathbf{k}(t),
	\end{equation}
	and, the total number of produced pairs per unit volume (in units of $m^3$) is obtained as
	\begin{equation}\label{eq:number density}
		N = \int \dfrac{d^3\mathbf{k}}{(2\pi)^3}f_\mathbf{k}(\infty).
	\end{equation}
	
	The ODE system \eqref{eq:DHW-10 ODEs} is solved using Runge-Kutta algorithms of various orders, including both adaptive and fixed-step schemes, over $t \in [-\eta\tau, \eta\tau]$, with sufficiently large $\eta (>0)$ to ensure the field vanishes at the integration boundaries. The implementation is benchmarked against Ref. \cite{olugh_pair_2019}, reproducing the published momentum distributions with excellent agreement. Momentum integration in Eq. \eqref{eq:number density} uses $k_{\mathrm{cutoff}}=10\,m$ and a uniform momentum grid spacing of $\Delta k=0.05\,m$. Convergence is verified by increasing the cutoff to $15\,m$, refining the grid to $\Delta k=0.02\,m$, and checking the stability against solver choice, time step, and integration boundaries.
%%%%%%%%%%%%%%%%%%%%%%%%%%%%%% %%%%%%%%%%%%%%%%%%%%%%%%%%%%%%%%%%%
\section{Polarization and chirp effects}\label{sec:polarization and chirp effects}
%%%%%%%%%%%%%%%%%%%%%%%%%%%%%%%%%%%%%%%%%%%%%%%%%%%%%%%%%%%%%%%%%
	In this section, we present our numerical results for $e^+e^-$ pair production under different polarization states of the electric field: (i) linear polarization ($\delta = 0$), (ii) elliptical polarization ($0<\delta<1$), and (iii) circular polarization ($\delta = 1$). We investigate the effects of symmetric and asymmetric quadratic frequency chirping and examine how the resulting modifications to the field evolution influence pair production across these polarization configurations. We present the momentum distributions in the $(k_x,k_y)$ plane with $k_z=0$, and work in the gauge $\mathbf{A}(+\infty)=0$, for which the canonical and kinetic momenta coincide at asymptotic times. Throughout this section, we set $\varphi=0$; the effects of a nonzero CEP are discussed separately in Sec.\ref{sec:CEP effects}.
	\begin{figure}[tbh]
		\centering
		\includegraphics[width=\columnwidth]{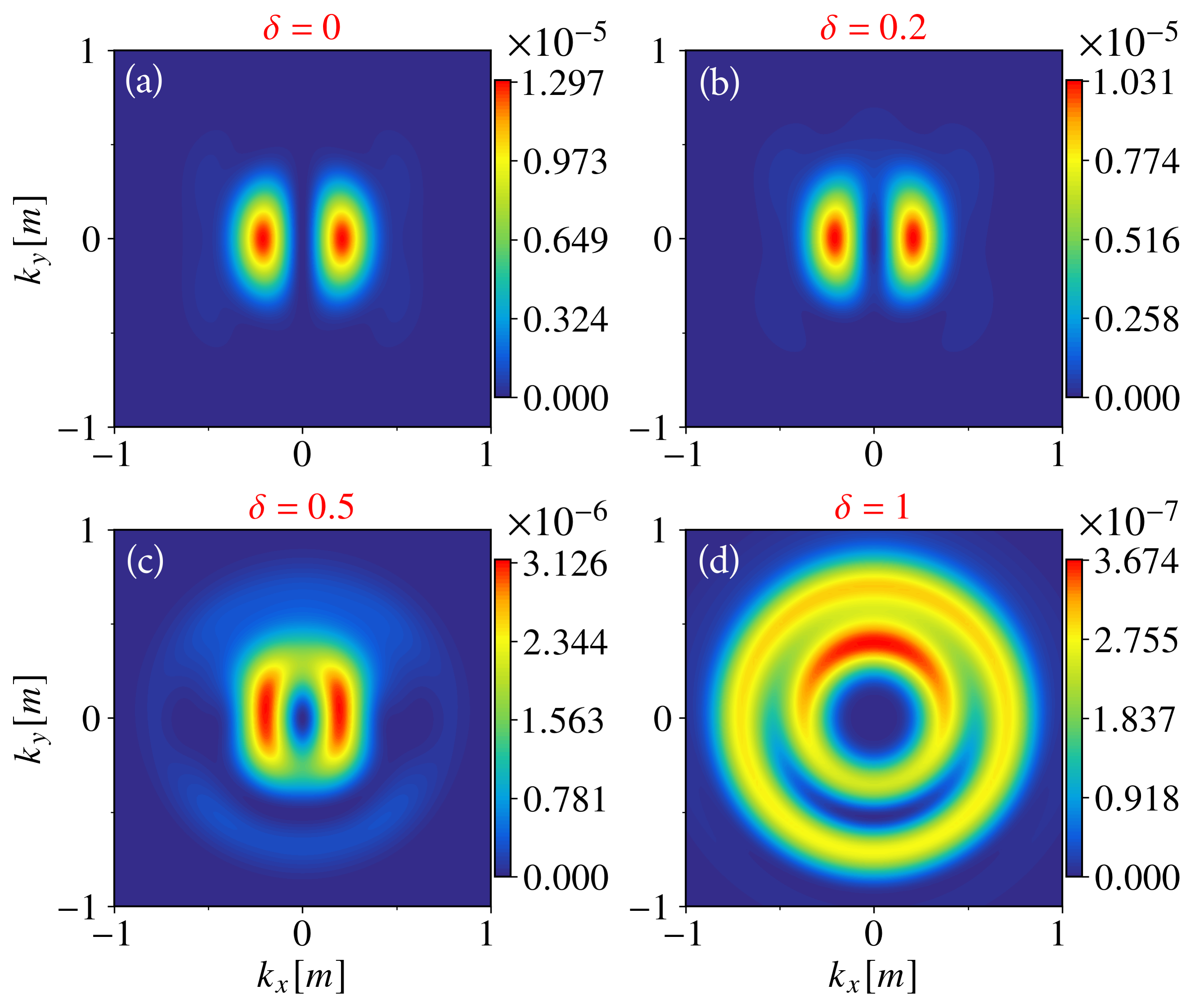}
		\caption{Momentum distribution of the created $e^+e^-$ pairs with $k_z = 0$. The panels show the variation with $\delta$ in a chirp free case ($b_1 = b_2 = 0$). All other field parameters are the same as those specified in Eq. \eqref{eq:field_parameters}.}
		\label{fig:MD_delta_all_chirp_free}
	\end{figure}
	\begin{figure*}[tbh]
		\centering
		\includegraphics[width=0.8\textwidth]{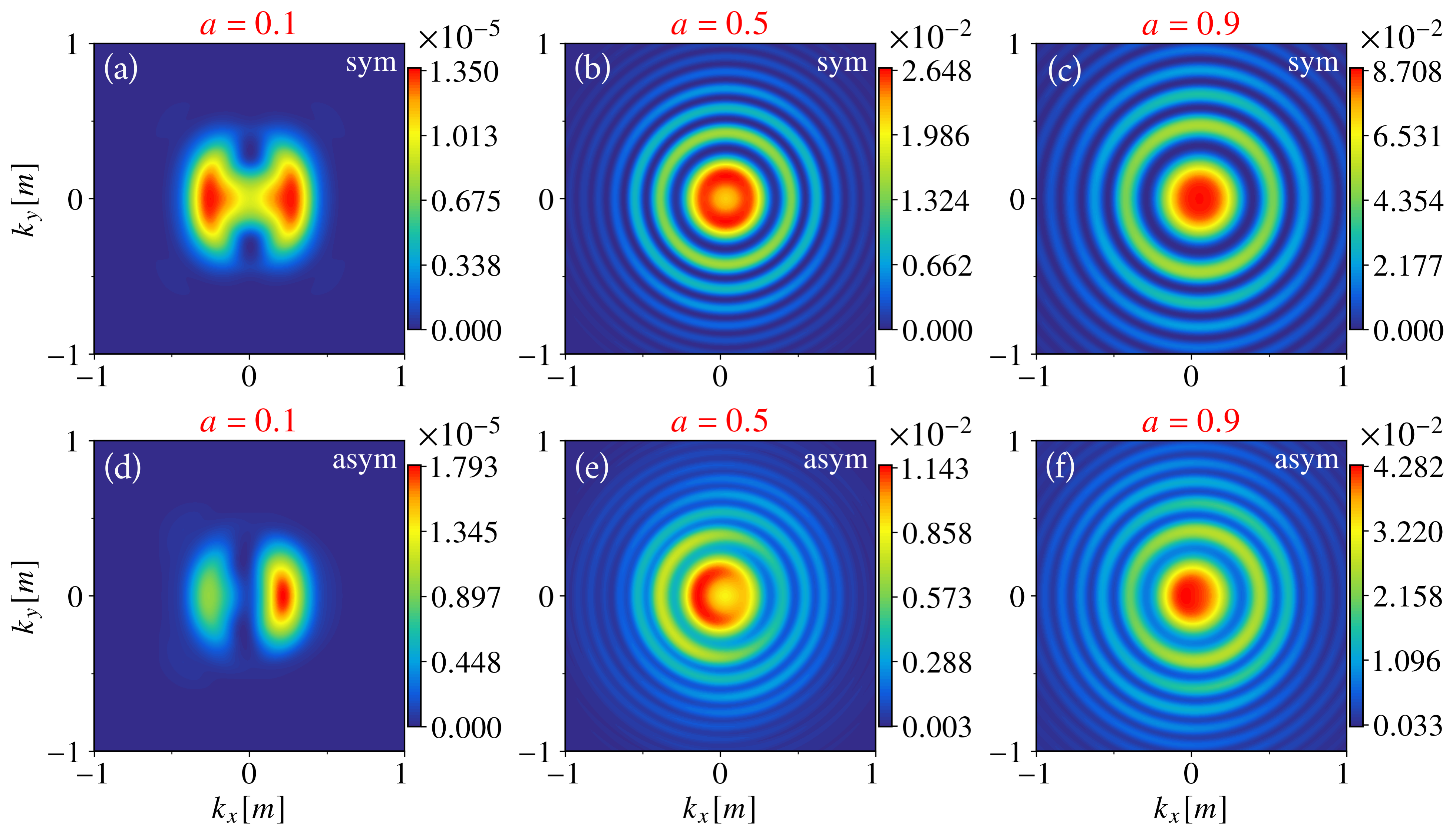}
		\caption{Momentum distribution of the created $e^+e^-$ pairs with $k_z=0$ for $\delta=0$. Panels (a)–(c) correspond to the \textit{symmetric} chirp, while panels (d)–(f) show the \textit{asymmetric} chirp. The columns (a,d), (b,e), and (c,f) correspond to the chirp-control parameter $a=0.1$, $0.5$, and $0.9$, respectively. The corresponding chirp parameters are given by $b_1=a\omega/(2\tau)$ and $b_2=a\omega/(2\tau^2)$. All other field parameters are the same as those specified in Eq. \eqref{eq:field_parameters}.}
		\label{fig:MD_delta_0_chirp_all}
	\end{figure*}
	
	Before examining the effects of chirping for each polarization state, we first establish the chirp-free ($b_1=b_2=0$) baseline by presenting the results for momentum distribution of the produced $e^+e^-$ pairs. The corresponding results are shown in Fig. \ref{fig:MD_delta_all_chirp_free}. The momentum spectra exhibits a clear dependence on $\delta$ with substantial changes in its shape, momentum-space extent, and peak values. For linear polarization ($\delta = 0$), the distribution consists of two well-localized peaks symmetrically positioned along the $k_x$ axis. As $\delta$ increases, the two-peak structure gradually broadens and develops a more complex pattern, with the emergence of additional features around the central momentum region. For $\delta=1$, corresponding to circular polarization, the momentum distribution develops two arc-like structures that approximately follow an annular shape, with pair production concentrated along a curved ridge surrounding a depleted central region. Concurrently, the overall momentum-space distribution spreads over a wider range of momenta. The peak magnitude, however, decreases markedly as $\delta$ increases, indicating a substantial suppression of pair production with increasing ellipticity for the chosen field parameters.

	Having established the chirp-free momentum distributions, we now investigate how symmetric and asymmetric quadratic frequency chirping modifies the momentum-space structure and peak occupation of the produced pairs for different polarization states. We consider three values of the chirp-control parameter, $a=0.1$, $0.5$, and $0.9$, corresponding to weak, intermediate, and strong chirping, respectively. The chirp parameters are defined as $b_1=a\omega/(2\tau)$ and $b_2=a\omega/(2\tau^2)$ for both symmetric and asymmetric chirping. 
	%--------------------------------------------------------------------------
	\subsection{Linearly polarized fields ($\delta = 0$)}\label{subsec:delta_0}
	%--------------------------------------------------------------------------
	For linear polarization ($\delta=0$), the momentum distributions exhibit a strong dependence on both the type and strength of the chirp, as shown in Fig.~\ref{fig:MD_delta_0_chirp_all}. For weak chirping ($a=0.1$), symmetric chirping largely preserves the characteristic two-lobed structure of the chirp-free distribution, with the peaks located symmetrically on either side of the origin along the $k_x$ axis. In contrast, asymmetric chirping breaks the symmetry between the two lobes, leading to a pronounced difference in their peak magnitudes. As the chirp strength increases, the momentum distributions undergo substantial changes, developing concentric, ring-like oscillatory structures around the central momentum region. These features become more pronounced at stronger chirping ($a = 0.9$), accompanied by a redistribution of the produced pairs across momentum space. The peak magnitude also responds differently to the two chirp types, with symmetric chirping producing substantially higher peaks than asymmetric chirping at stronger chirp strengths. Overall, these results demonstrate that quadratic frequency chirping can significantly reshape the momentum distributions and modify the pair-production yield, while asymmetric chirping can additionally induce a pronounced momentum-space imbalance.
	%--------------------------------------------------------------------------
	\subsection{Elliptically polarized fields ($0 < \delta < 1$)}\label{subsec:delta_0.2,0.5}
	%--------------------------------------------------------------------------
	\begin{figure*}[tbh]
		\centering
		\includegraphics[width=0.8\textwidth]{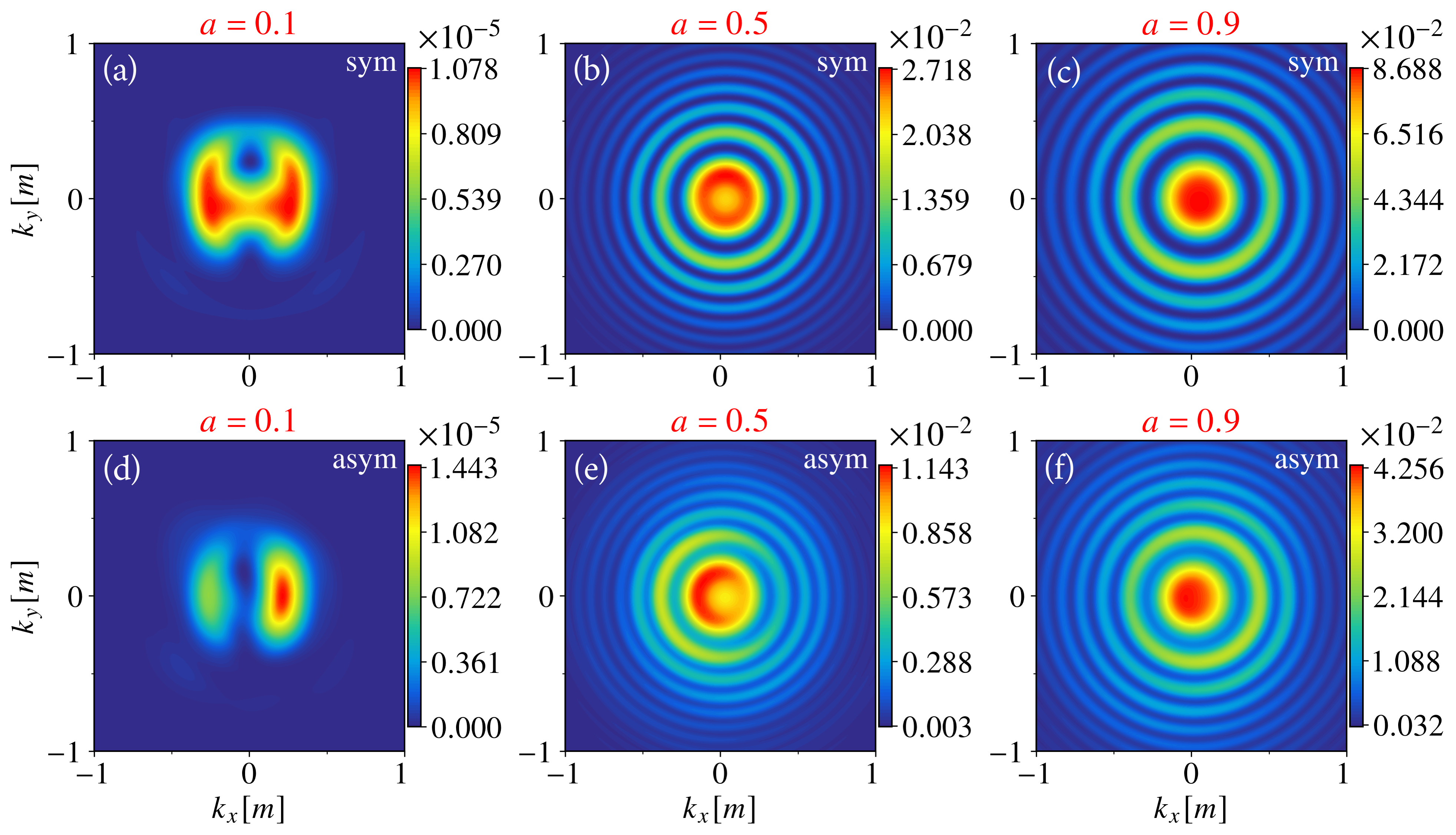}
		\caption{Momentum distribution of the created $e^+e^-$ pairs with $k_z=0$ for $\delta=0.2$. Panels (a)–(c) correspond to the \textit{symmetric} chirp, while panels (d)–(f) show the \textit{asymmetric} chirp. The columns (a,d), (b,e), and (c,f) correspond to the chirp-control parameter $a=0.1$, $0.5$, and $0.9$, respectively. The corresponding chirp parameters are given by $b_1=a\omega/(2\tau)$ and $b_2=a\omega/(2\tau^2)$. All other field parameters are the same as those specified in Eq. \eqref{eq:field_parameters}.}
		\label{fig:MD_delta_0.2_chirp_all}
	\end{figure*}
	\begin{figure*}[tbh]
		\centering
		\includegraphics[width=0.8\textwidth]{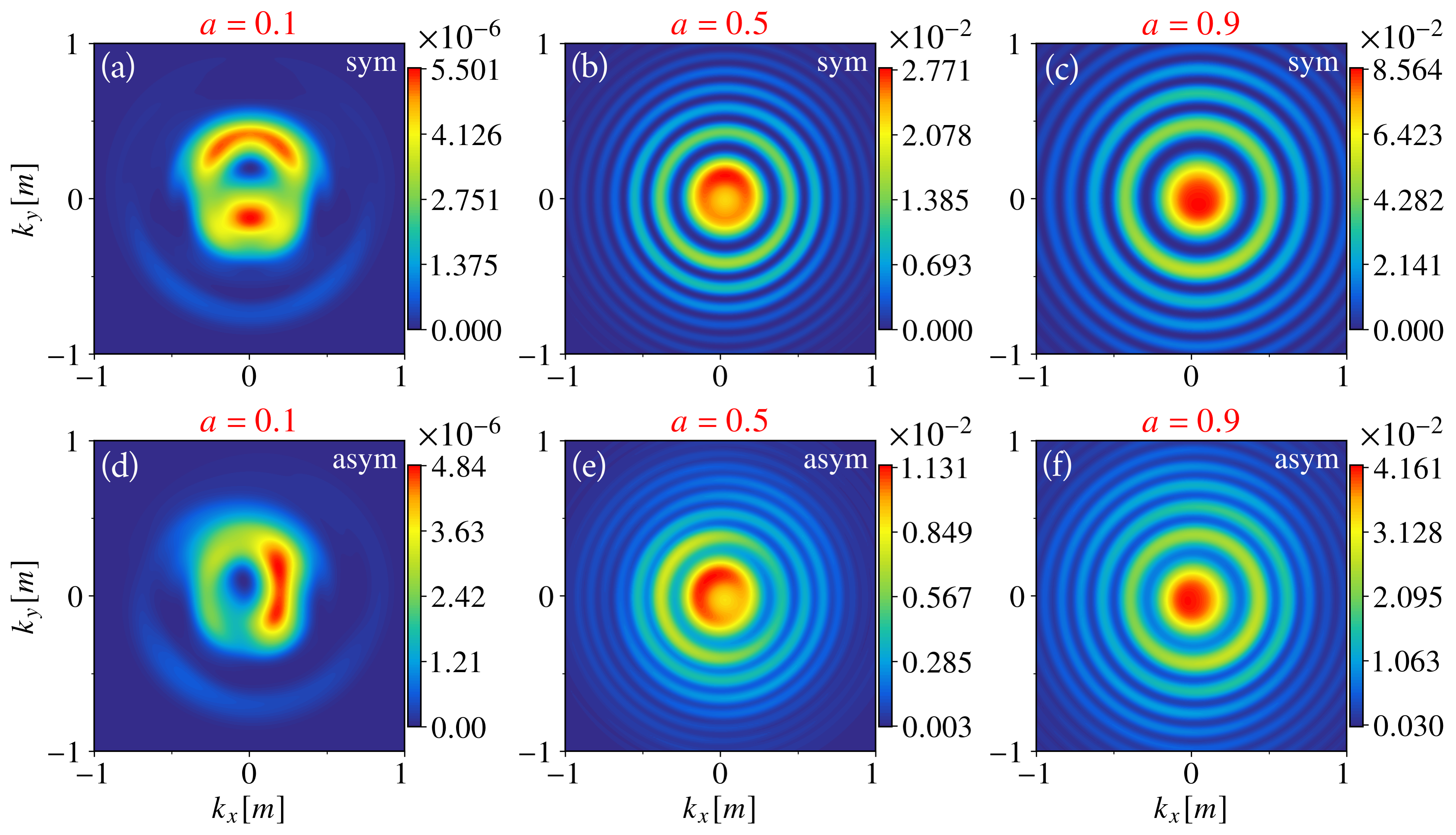}
		\caption{Momentum distribution of the created $e^+e^-$ pairs with $k_z=0$ for $\delta=0.5$. Panels (a)–(c) correspond to the \textit{symmetric} chirp, while panels (d)–(f) show the \textit{asymmetric} chirp. The columns (a,d), (b,e), and (c,f) correspond to the chirp-control parameter $a=0.1$, $0.5$, and $0.9$, respectively. The corresponding chirp parameters are given by $b_1=a\omega/(2\tau)$ and $b_2=a\omega/(2\tau^2)$. All other field parameters are the same as those specified in Eq. \eqref{eq:field_parameters}.}
		\label{fig:MD_delta_0.5_chirp_all}
	\end{figure*}
	\begin{figure*}[tbh]
		\centering
		\includegraphics[width=0.8\textwidth]{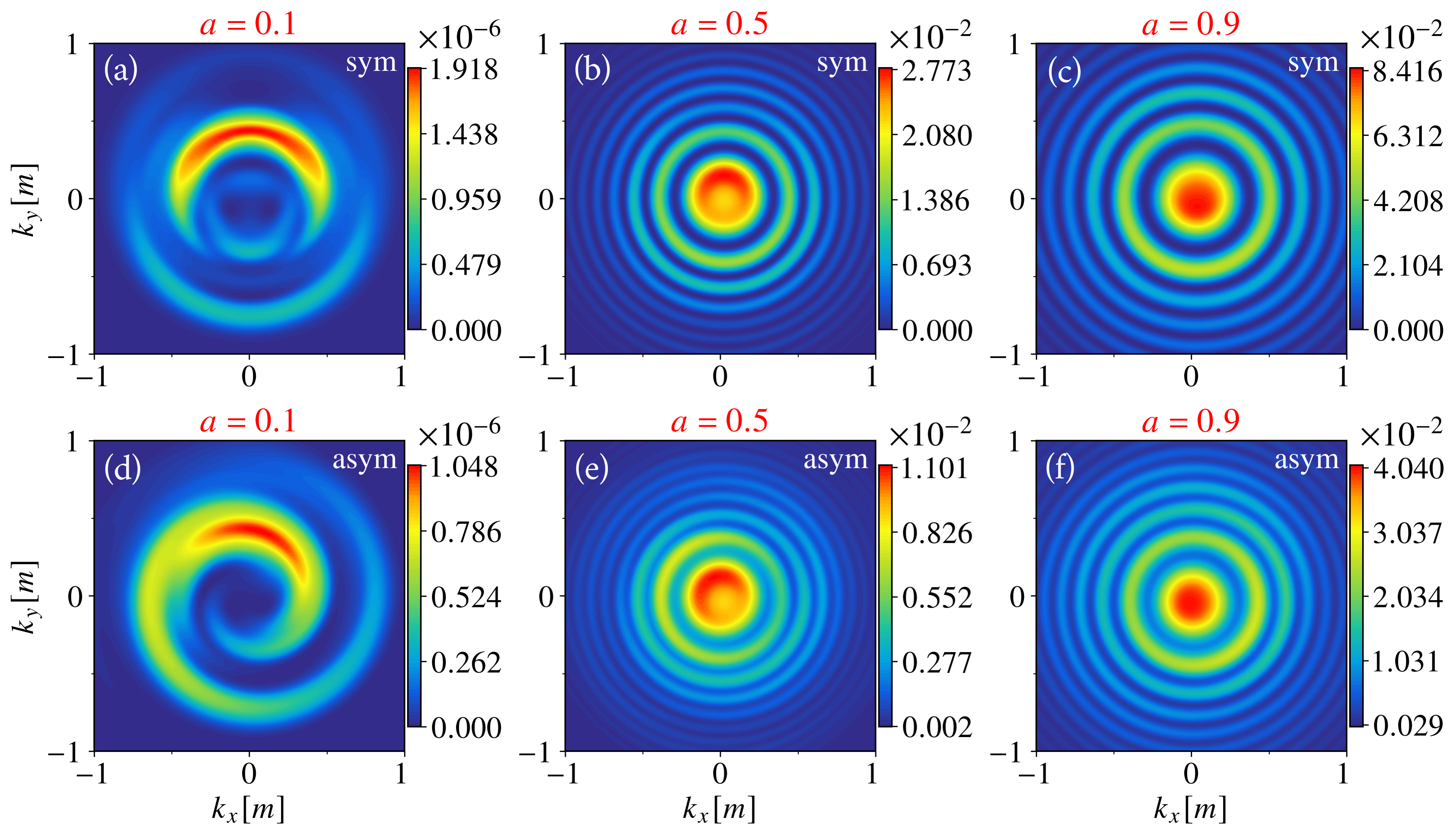}
		\caption{Momentum distribution of the created $e^+e^-$ pairs with $k_z=0$ for $\delta=1$. Panels (a)–(c) correspond to the \textit{symmetric} chirp, while panels (d)–(f) show the \textit{asymmetric} chirp. The columns (a,d), (b,e), and (c,f) correspond to the chirp-control parameter $a=0.1$, $0.5$, and $0.9$, respectively. The corresponding chirp parameters are given by $b_1=a\omega/(2\tau)$ and $b_2=a\omega/(2\tau^2)$. All other field parameters are the same as those specified in Eq. \eqref{eq:field_parameters}.}
		\label{fig:MD_delta_1_chirp_all}
	\end{figure*}
	For elliptical polarization, corresponding to $\delta=0.2$ and $\delta=0.5$, the momentum distributions exhibit a pronounced dependence on both the chirp strength and the type of chirping, as shown in Figs. \ref{fig:MD_delta_0.2_chirp_all} and \ref{fig:MD_delta_0.5_chirp_all}. At weak chirping ($a=0.1$), the distributions retain a nonuniform, angularly structured pattern rather than the two-lobed structure characteristic of the linearly polarized case. For $\delta=0.2$, symmetric chirping [Fig. \ref{fig:MD_delta_0.2_chirp_all}(a)] produces a curved, symmetric butterfly-like distribution with two prominent lobes, whereas asymmetric chirping [Fig. \ref{fig:MD_delta_0.2_chirp_all}(d)] yields two distinct peaks with unequal magnitudes. At $\delta=0.5$, the distributions become more extended in the transverse momentum direction: symmetric chirping [Fig. \ref{fig:MD_delta_0.5_chirp_all}(a)] produces two prominent regions above and below the origin, while asymmetric chirping [Fig. \ref{fig:MD_delta_0.2_chirp_all}(d)] generates spiral-like structure. These patterns differ from the chirp-free distributions [Figs. \ref{fig:MD_delta_all_chirp_free}(b) and \ref{fig:MD_delta_all_chirp_free}(c)], demonstrating that even weak chirping can substantially rearrange the momentum-space structure for elliptically polarized fields.
	
	As the chirp strength increases to $a=0.5$ and $a=0.9$, both chirp types produce well-defined concentric oscillatory rings centered around the origin, accompanied by a strong central maximum. This evolution is associated with the increase in the effective frequency $\omega_{\mathrm{eff}}(t)$ as the chirp strength grows, which modifies the energy supplied by the field and enhances pair production. The resulting momentum spectra exhibit increasingly pronounced interference patterns, manifested as concentric oscillatory rings. With increasing chirp strength, the radial organization of the momentum distributions becomes more prominent, and the overall structures for the two ellipticities become increasingly similar. This behavior is qualitatively consistent with the trend observed for linear polarization, indicating that stronger quadratic chirping can dominate the momentum-space structure despite changes in the field polarization. However, the peak magnitude depends on the chirp type: at moderate and strong chirping, symmetric chirping produces substantially higher peaks than asymmetric chirping for both values of $\delta$. At weak chirping, the relative peak heights are more sensitive to the ellipticity and chirp type. Compared with the chirp-free case, strong chirping therefore produces a marked enhancement of the peak occupation and transforms the polarization-dependent distributions into pronounced ring-like patterns. 
	%--------------------------------------------------------------------------
	\subsection{Circularly polarized fields ($\delta = 1$)}\label{subsec:delta_1}
	%--------------------------------------------------------------------------
	For circular polarization ($\delta=1$), the momentum distributions exhibit a pronounced transition from an arc-like structure at weak chirping to concentric oscillatory rings at stronger chirping, as shown in Fig.~\ref{fig:MD_delta_1_chirp_all}. For $a=0.1$, symmetric chirping preserves the approximately annular structure of the chirp-free distribution (cf. Figs. \ref{fig:MD_delta_all_chirp_free}(d) and \ref{fig:MD_delta_1_chirp_all}(a)), with pair production concentrated along a curved ridge and a depleted central region. Asymmetric chirping, by contrast, produces a spiral-like pattern with a more evident angular asymmetry. In both cases, the peak occupation remains relatively small. Increasing the chirp strength to $a=0.5$ transforms the distributions into a series of concentric rings surrounding a prominent central maximum. This structure becomes more pronounced for $a=0.9$, with the central peak increasing further under both chirp types. However, symmetric chirping produces a higher peak magnitude than asymmetric chirping at both moderate and strong chirp strengths similar to previous cases.
	\begin{figure*}[tbh]
		\centering
		\includegraphics[width=\textwidth]{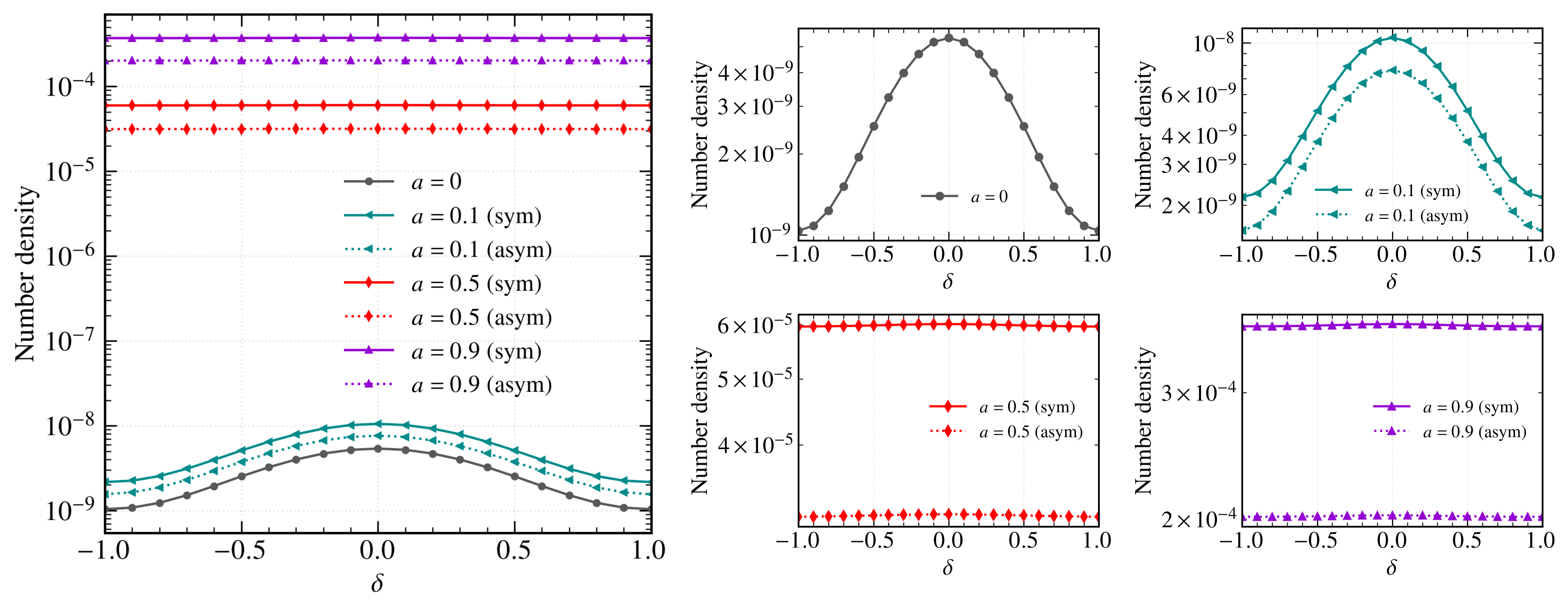}
		\caption{Number density (in units of $m^3$) of created particles as a function of the field polarization $\delta$ for various chirp-control parameter $a$. The corresponding chirp parameters are given by $b_1=a\omega/(2\tau)$ and $b_2=a\omega/(2\tau^2)$. The other field parameters are fixed to the values specified in Eq.~\eqref{eq:field_parameters}.}
		\label{fig:ND_delta_all_phi_0}
	\end{figure*}
	
	Compared with the chirp-free circularly polarized case [Fig. \ref{fig:MD_delta_all_chirp_free}(d)], in which the distribution is predominantly arc-like and the central region is depleted, stronger chirping substantially reorganizes the momentum-space structure and enhances the peak occupation. The low-chirp distributions retain clear polarization-dependent differences: linear polarization exhibits a characteristic two-lobed structure, elliptical polarization develops more complex curved and partially ring-like patterns, and circular polarization displays an annular or spiral-like structure. Thus, while stronger chirping drives the distributions toward a common ring-like morphology, the initial momentum-space structure and the relative peak magnitudes remain sensitive to both polarization and chirp type.

	We now examine the total number density of the produced particles as a function of the polarization parameter $\delta$ for different chirp strengths, as shown in Fig.~\ref{fig:ND_delta_all_phi_0}. In the absence of chirping ($a=0$), the number density exhibits a clear dependence on polarization, attaining its maximum near $\delta=0$ and decreasing as the field approaches circular polarization ($\delta=1$). This indicates that, for the chirp-free field, the particle yield is largest for linear polarization and decreases with increasing ellipticity.
	
	For weak chirping ($a=0.1$), the number density increases relative to the chirp-free case, while retaining a similar dependence on $\delta$. It reaches its maximum near linear polarization and decreases toward circular polarization for both symmetric and asymmetric chirping. However, symmetric chirping consistently produces a higher number density than asymmetric chirping across the polarization range. Thus, at weak chirp strength, the overall yield remains sensitive to the field polarization, although chirping enhances particle production compared with the chirp-free case.
	
	A markedly different behavior emerges at stronger chirping. For $a=0.5$, the number density increases substantially for both chirp types, with symmetric chirping yielding a considerably larger particle population than asymmetric chirping. Moreover, the number density becomes weakly dependent on $\delta$, indicating that the polarization dependence observed at weaker chirping is strongly suppressed. This trend persists at $a=0.9$, where the number density increases further and remains approximately constant across the polarization range, with symmetric chirping continuing to produce a higher yield than asymmetric chirping. The suppression of the polarization dependence of the total number density at stronger chirping is consistent with previous findings for linearly chirped fields reported in Refs. \cite{jangir_dynamically_2026, olugh_pair_2019}.
	
	Overall, Fig.~\ref{fig:ND_delta_all_phi_0} shows that increasing the chirp strength enhances the total particle yield while progressively reducing its dependence on polarization. Whereas polarization strongly influences pair production in the chirp-free and weakly chirped regimes, its effect on the integrated yield becomes negligible at stronger chirping. This behavior suggests that frequency modulation increasingly dominates the total production yield, although polarization continues to shape the momentum distributions. Furthermore, symmetric chirping consistently produces a higher number density than asymmetric chirping, highlighting the role of the temporal structure of the frequency modulation in determining the overall pair yield.
	\begin{figure*}[tbh]
		\centering
		\includegraphics[width=\textwidth]{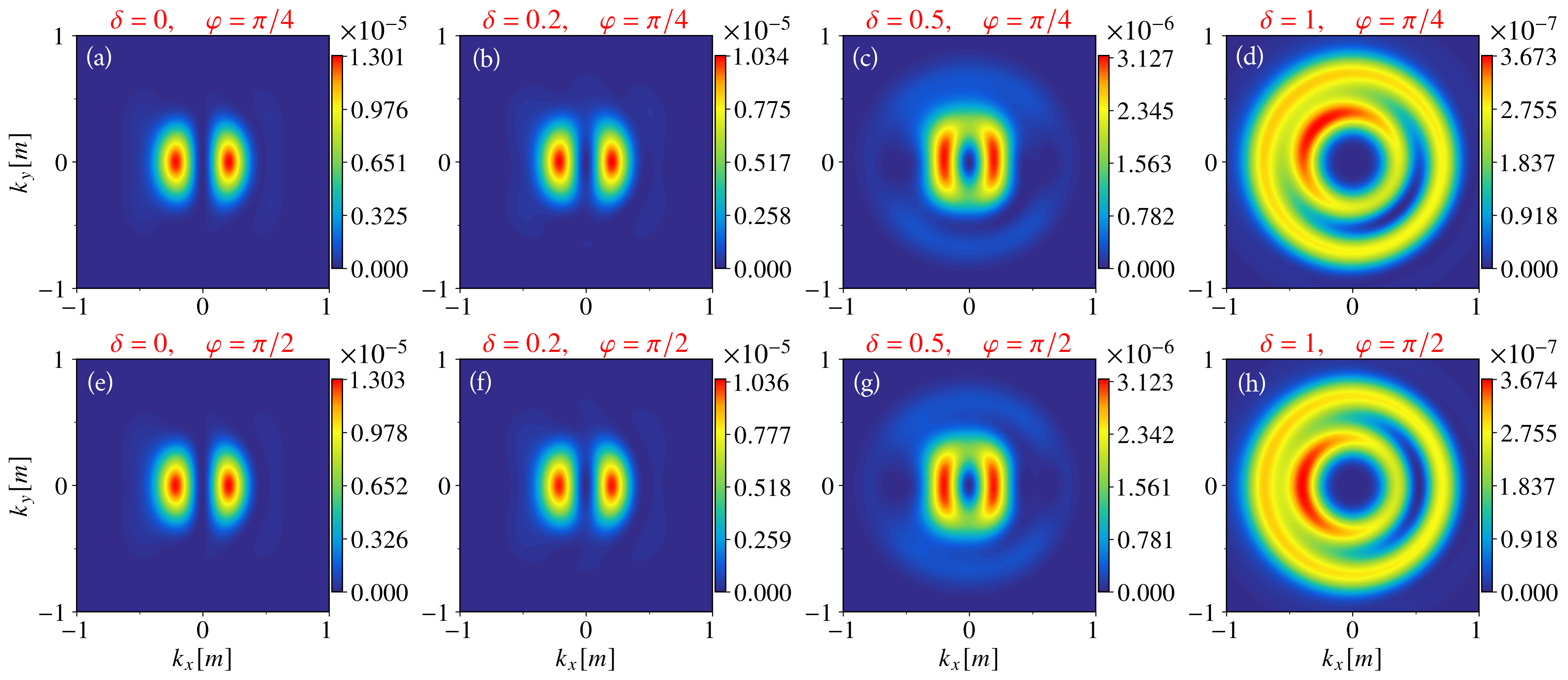}
		\caption{Momentum distribution of the created $e^+e^-$ pairs with $k_z=0$ in the chirp-free case ($b_1 = b_2 = 0$). Panels (a)–(d) correspond to the CEP $\varphi=\pi/4$, while panels (e)–(h) correspond to $\varphi=\pi/2$. The columns (a,e), (b,f), (c,g), and (d,h) show the results for $\delta=0$, $0.2$, $0.5$, and $1$, respectively. All other field parameters are the same as those specified in Eq. \eqref{eq:field_parameters}.}
		\label{fig:MD_delta_all_chirp_free_phi_all}
	\end{figure*}
%%%%%%%%%%%%%%%%%%%%%%%%%%%%%% %%%%%%%%%%%%%%%%%%%%%%%%%%%%%%%%%%%
\section{Carrier-envelope phase effects}\label{sec:CEP effects}
%%%%%%%%%%%%%%%%%%%%%%%%%%%%%%%%%%%%%%%%%%%%%%%%%%%%%%%%%%%%%%%%%
	So far, we have considered the CEP to be $\varphi=0$. In this section, we investigate the effects of a nonzero CEP on $e^+e^-$ pair production under different polarization states and frequency-chirping conditions. We examine how variations in the CEP modify the $e^+e^-$ pair production, highlighting its interplay with polarization and chirping.

	We first investigate the effects of a nonzero CEP in the absence of frequency chirping, with $b_1=b_2=0$. Figure~\ref{fig:MD_delta_all_chirp_free_phi_all} shows the momentum distributions for $\varphi=\pi/4$ and $\varphi=\pi/2$ at different polarization states, $\delta=0$, $0.2$, $0.5$, and $1$. Comparing these results with the chirp-free distributions at $\varphi=0$ [Fig. \ref{fig:MD_delta_all_chirp_free}], we find that the influence of the CEP depends on the polarization state of the electric field.
	
	For linear polarization ($\delta=0$), the momentum distribution retains its characteristic two-peak structure, with the maxima located on opposite sides of the origin along the $k_x$ axis. The peak occupation is approximately $1.30\times10^{-5}$ for both $\varphi=\pi/4$ and $\varphi=\pi/2$. The CEP produces no appreciable change in the maximum occupation between these two phase values, and the overall two-lobed structure remains similar to that obtained for $\varphi=0$. A comparable two-peak structure persists for weak ellipticity ($\delta=0.2$), although the peak occupation decreases to approximately $1.03\times10^{-5}$. The distributions remain concentrated around two distinct momentum regions, with only modest changes in their detailed profiles as the CEP varies.
	
	For $\delta=0.5$, the momentum distribution develops a more complex structure, consisting of two pronounced inner peaks surrounded by a diffuse, approximately ring-shaped feature. The maximum occupation decreases further to approximately $3.13\times10^{-6}$ for both CEP values. This reduction, together with the broadening of the distribution in momentum space, reflects the strong dependence of pair production on the polarization state. Compared with the corresponding chirp-free result at $\varphi=0$, the overall morphology remains consistent with the transition from a two-peak distribution toward a more extended, ring-like pattern as the ellipticity increases.
	
	A distinct behavior occurs for circular polarization ($\delta=1$). The momentum distribution exhibits concentric ring-like structures surrounding a depleted central region, with a maximum occupation of approximately $3.67\times10^{-7}$ for both $\varphi=\pi/4$ and $\varphi=\pi/2$. Importantly, the changes in the angular orientation of this distribution with the CEP arise directly from the form of the electric field \eqref{eq:field_model} adopted in this work. For circular polarization, varying $\varphi$ effectively rotates the momentum distribution in the $(k_x,k_y)$ plane rather than fundamentally altering its radial structure. Consequently, the CEP-dependent patterns for $\delta=1$ should be interpreted primarily as rotations of the same underlying distribution.
	\begin{figure*}[tbh]
		\centering
		\includegraphics[width=\textwidth]{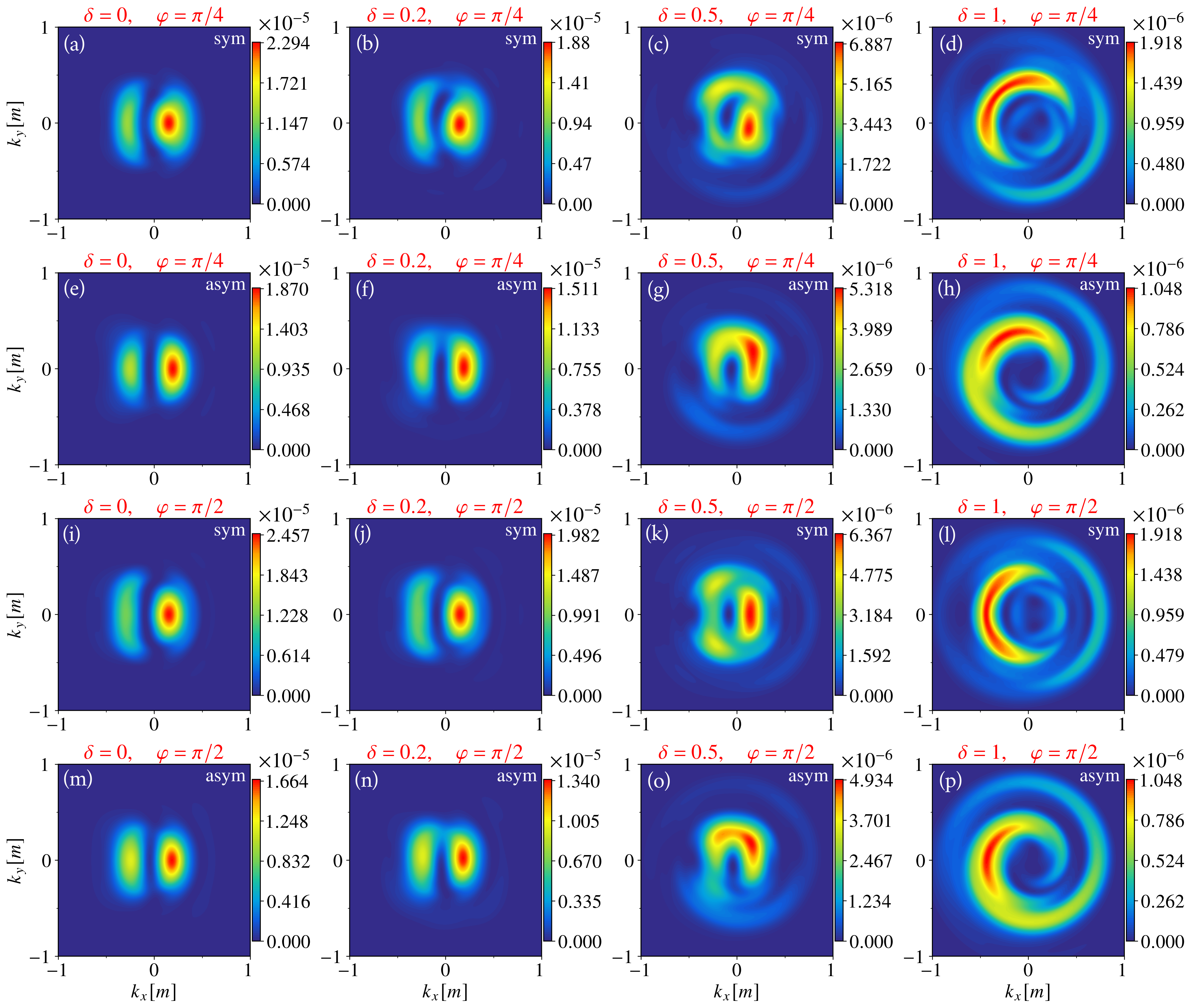}
		\caption{Momentum distribution of the created $e^+e^-$ pairs with $k_z=0$ for different polarization, CEP, and chirp configurations. Panels (a)–(d) and (e)–(h) correspond to $\varphi=\pi/4$ for the \textit{symmetric} and \textit{asymmetric} chirps, respectively, while panels (i)–(l) and (m)–(p) correspond to $\varphi=\pi/2$ for the \textit{symmetric} and \textit{asymmetric} chirps, respectively. The columns (a,m), (b,n), (c,o), and (d,p) correspond to $\delta=0$, $0.2$, $0.5$, and $1$, respectively. The chirp parameters are given by $b_1=a\omega/(2\tau)$ and $b_2=a\omega/(2\tau^2)$, with the chirp-control parameter fixed at $a=0.1$. All other field parameters are the same as those specified in Eq. \eqref{eq:field_parameters}.}
		\label{fig:MD_delta_all_a_0.1_phi_all}
	\end{figure*}
	
	We now investigate the effects of a nonzero CEP in the presence of symmetric and asymmetric frequency chirping. Figure \ref{fig:MD_delta_all_a_0.1_phi_all} presents the momentum distributions for $\varphi=\pi/4$ and $\pi/2$ with the chirp-control parameter fixed at $a=0.1$. The results show that the CEP modifies the momentum-space structure in a polarization-dependent manner, while the response of the peak occupation depends on both the chirp type and the polarization state. Compared with the chirp-free distributions [Fig. \ref{fig:MD_delta_all_chirp_free_phi_all}], weak chirping introduces additional changes in the shape and relative prominence of the momentum-space features.
	
	For linear polarization ($\delta=0$), the distributions retain the characteristic two-lobed structure, with the dominant peaks located on opposite sides of the origin along the $k_x$ direction. Symmetric chirping produces a more pronounced dominant peak, accompanied by a weaker neighboring lobe, whereas asymmetric chirping yields a lower maximum occupation. Changing the CEP from $\pi/4$ to $\pi/2$ modifies the relative prominence of the two lobes and slightly changes the peak occupation, but does not fundamentally alter the overall structure. Thus, for linear polarization, the CEP and weak chirping primarily affect the relative peak heights and detailed shape of the distribution rather than its characteristic two-peak morphology.
	
	A similar overall structure persists for weak ellipticity ($\delta=0.2$), although the momentum distributions become somewhat broader and less symmetric in their local features. The two dominant momentum regions remain clearly identifiable for both CEP values and chirp types. As in the linearly polarized case, symmetric chirping produces higher peak occupations than asymmetric chirping. The change in CEP modifies the peak prominence and detailed distribution, but the overall morphology remains similar. These results indicate that weak ellipticity does not qualitatively eliminate the two-lobed structure, although it changes the momentum-space distribution and reduces the peak occupation relative to linear polarization.
	
	For stronger ellipticity ($\delta=0.5$), the momentum distributions exhibit more intricate structures. In addition to the prominent central momentum-space features, curved ridges and diffuse ring-like patterns become visible. Their arrangement varies with the CEP and the type of chirping: symmetric chirping produces a pronounced curved structure surrounding the central region, whereas asymmetric chirping leads to a differently shaped and more angularly asymmetric distribution. The maximum occupation is lower than for $\delta=0$ and $\delta=0.2$, and asymmetric chirping again yields a smaller peak than symmetric chirping. The differences between $\varphi=\pi/4$ and $\pi/2$ are particularly apparent in the arrangement of the curved features, demonstrating that the CEP can substantially modify the detailed momentum-space morphology even when the overall occupation remains of a similar order.
	
	For circular polarization ($\delta=1$), the momentum distributions exhibit ring-like structures with a depleted central region. The angular orientation of these structures changes with the CEP, while their overall radial organization remains similar. This behavior follows directly from the form of the electric field adopted in this work: for circular polarization, varying the CEP effectively rotates the momentum distribution in the $(k_x,k_y)$ plane. Accordingly, the changes between $\varphi=\pi/4$ and $\pi/2$ primarily reflect a rotation of the distribution rather than a fundamental change in its structure. Symmetric chirping produces a more pronounced peak occupation than asymmetric chirping, although the latter retains the characteristic ring-like pattern with a different angular distribution of the produced pairs.
	
	Overall, the results demonstrate that the CEP primarily modifies the momentum-space distributions, with its effects depending on the polarization state and the presence of frequency chirping. In both chirp-free and chirped fields, variations in the CEP alter the shapes, orientations, and relative peak occupations of the distributions, while their characteristic structures remain largely governed by the field polarization and chirp parameters. For circular polarization, the CEP dependence manifests primarily as a rotation of the momentum distribution in the $(k_x,k_y)$ plane, as dictated by the adopted field configuration. In contrast, linear and elliptical polarization exhibit more noticeable changes in the relative peak heights and detailed momentum-space structures. Since the total number density exhibits no significant variation with the CEP over the parameter range considered, the corresponding results are omitted for brevity.
%%%%%%%%%%%%%%%%%%%%%%%%%%%%%%%%%%%%%%%%%%%%%%%%%%%%%%%%%%%%%%%%%
\section{Conclusion and Discussion}\label{sec:Conclusion}
%%%%%%%%%%%%%%%%%%%%%%%%%%%%%%%%%%%%%%%%%%%%%%%%%%%%%%%%%%%%%%%%%
In this work, we investigate electron–positron pair production in spatially homogeneous, time-dependent electric fields with different polarization states using the Dirac–Heisenberg–Wigner formalism. We systematically examine the effects of symmetric and asymmetric quadratic frequency chirping and the carrier-envelope phase, focusing on their influence on the momentum distributions and total number density of the produced pairs. Our results demonstrate that the interplay between polarization and frequency modulation provides an effective means of modifying both the momentum-space structure and the overall yield of pair production.

The momentum distributions exhibit a strong dependence on polarization. Without chirping, the two-peak structure for linear polarization gradually evolves into extended, arc-like, and nearly annular structures as the field approaches circular polarization. Weak chirping largely preserves these characteristic features, with modest changes in their shapes and enhanced peak occupations. In contrast, strong chirping produces pronounced oscillatory and ring-like structures that persist across all polarization states.

The total number density increases with chirp strength, with symmetric quadratic chirping generally yielding more pairs than asymmetric chirping. In the weakly chirped regime, the yield depends strongly on polarization, reaching its maximum for linear polarization. This dependence diminishes rapidly with increasing chirp strength, becoming negligible under strong chirping, for which the yield is nearly independent of polarization.

We further investigate the effects of a nonzero CEP in both chirp-free and chirped fields. The CEP modifies the momentum distributions by changing their orientation, shape, and relative peak occupations, with the extent of these changes depending on the polarization state. For circular polarization, varying the CEP primarily rotates the momentum distribution in the $(k_x,k_y)$ plane, as dictated by the adopted field configuration. For linear and elliptical polarization, the CEP can modify the detailed momentum-space structure while preserving its principal characteristics. 

In summary, our findings establish that polarization, quadratic frequency chirping, and CEP influence pair production in complementary ways. Polarization determines the underlying momentum-space morphology, chirping can substantially reshape the distributions and enhance the total yield, and the CEP primarily affects the detailed momentum-space structure. This study provides insight into the control of Schwinger pair production in temporally structured electric fields and emphasizes the importance of considering both the momentum-resolved distributions and the integrated particle yield when assessing the effects of external-field parameters.
%%%%%%%%%%%%%%%%%%%%%%%%%%%%%%%%%%%%%%%%%%%%%%%%%%%%%%%%%%%%%%%%%
\bibliographystyle{apsrev4-2}
\bibliography{references}

@article{schneider_dynamically_2016,
	title = {Dynamically assisted {Sauter}-{Schwinger} effect in inhomogeneous electric fields},
	volume = {2016},
	issn = {1029-8479},
	url = {https://doi.org/10.1007/JHEP02(2016)164},
	doi = {10.1007/JHEP02(2016)164},
	number = {2},
	journal = {Journal of High Energy Physics},
	author = {Schneider, Christian and Schützhold, Ralf},
	month = feb,
	year = {2016},
	pages = {164},
}

@article{jansen_strong_2016,
	title = {Strong-field Breit-Wheeler pair production in short laser pulses: Identifying multiphoton interference and carrier-envelope-phase effects},
	author = {Jansen, Martin J. A. and M{\"u}ller, Carsten},
	journal = {Phys. Rev. D},
	volume = {93},
	issue = {5},
	pages = {053011},
	numpages = {13},
	year = {2016},
	month = {Mar},
	publisher = {American Physical Society},
	doi = {10.1103/PhysRevD.93.053011},
	url = {https://link.aps.org/doi/10.1103/PhysRevD.93.053011}
}

@book{fradkin_quantum_1991,
	title     = {Quantum Electrodynamics with Unstable Vacuum},
	author    = {Fradkin, Efim Samoilovich and Gitman, Dmitriy M. and Shvartsman, Sh. M.},
	year      = {1991},
	publisher = {Springer},
	address   = {Berlin, Heidelberg},
	url = {https://www.osti.gov/etdeweb/biblio/6972673}
}

@article{krajewska_phase_2012,
	title = {Phase effects in laser-induced electron-positron pair creation},
	author = {Krajewska, K. and Kami{\ifmmode \acute{n}\else \'{n}\fi{}}ski, J. Z.},
	journal = {Phys. Rev. A},
	volume = {85},
	issue = {4},
	pages = {043404},
	numpages = {12},
	year = {2012},
	month = {Apr},
	publisher = {American Physical Society},
	doi = {10.1103/PhysRevA.85.043404},
	url = {https://link.aps.org/doi/10.1103/PhysRevA.85.043404}
}

@article{birula_phase_1991,
	title = {Phase-space structure of the Dirac vacuum},
	author = {Bialynicki-Birula, Iwo and G{\'o}rnicki, Pawel and Rafelski, Johann},
	journal = {Phys. Rev. D},
	volume = {44},
	issue = {6},
	pages = {1825--1835},
	numpages = {0},
	year = {1991},
	month = {Sep},
	publisher = {American Physical Society},
	doi = {10.1103/PhysRevD.44.1825},
	url = {https://link.aps.org/doi/10.1103/PhysRevD.44.1825}
}

@article{bell_pissibility_2008,
	title = {Possibility of Prolific Pair Production with High-Power Lasers},
	author = {Bell, A. R. and Kirk, John G.},
	journal = {Phys. Rev. Lett.},
	volume = {101},
	issue = {20},
	pages = {200403},
	numpages = {4},
	year = {2008},
	month = {Nov},
	publisher = {American Physical Society},
	doi = {10.1103/PhysRevLett.101.200403},
	url = {https://link.aps.org/doi/10.1103/PhysRevLett.101.200403}
}

@article{banerjee_electron_2018,
	title = {Electron–positron pair creation by counterpropagating laser pulses: role of carrier envelope phase},
	volume = {72},
	issn = {1434-6079},
	url = {https://doi.org/10.1140/epjd/e2017-80399-7},
	doi = {10.1140/epjd/e2017-80399-7},
	number = {1},
	journal = {The European Physical Journal D},
	author = {Banerjee, Chitradip and Singh, Manoranjan P.},
	year = {2018},
	pages = {4},
}

@article{bras_two_2020,
	title = {Two-color phase-of-the-phase spectroscopy applied to nonperturbative electron-positron pair production in strong oscillating electric fields},
	author = {Bra{\ss{}}, J. and Milbradt, R. and Villalba-Ch{\'a}vez, S. and Paulus, G. G. and M{\"u}ller, C.},
	journal = {Phys. Rev. A},
	volume = {101},
	issue = {4},
	pages = {043401},
	numpages = {8},
	year = {2020},
	month = {Apr},
	publisher = {American Physical Society},
	doi = {10.1103/PhysRevA.101.043401},
	url = {https://link.aps.org/doi/10.1103/PhysRevA.101.043401}
}

@article{esposti_worldline_2023,
	title = {Worldline instantons for the momentum spectrum of Schwinger pair production in spacetime dependent fields},
	author = {Degli Esposti, Gianluca and Torgrimsson, Greger},
	journal = {Phys. Rev. D},
	volume = {107},
	issue = {5},
	pages = {056019},
	numpages = {17},
	year = {2023},
	month = {Mar},
	publisher = {American Physical Society},
	doi = {10.1103/PhysRevD.107.056019},
	url = {https://link.aps.org/doi/10.1103/PhysRevD.107.056019}
}

@misc{jangir_effects_2026,
	title={Effects of super-Gaussian pulse shape and relative phase on dynamically assisted pair production in spatially inhomogeneous electric fields with frequency chirping}, 
	author={Abhinav Jangir and Anees Ahmed},
	year={2026},
	eprint={2609.26013},
	archivePrefix={arXiv},
	primaryClass={hep-ph},
	url={https://arxiv.org/abs/2609.26013}, 
}

@misc{chen_electron_2026,
	title={Electron-positron pair production in spatially inhomogeneous electric fields with quadratically symmetric chirp}, 
	author={Le Le Chen and Xiao-Ting Xu and Rong-An-Tang and Xue-Ren Hong and Lie-Juan Li and Bai-Song Xie},
	year={2026},
	eprint={2603.15196},
	archivePrefix={arXiv},
	primaryClass={hep-ph},
	url={https://arxiv.org/abs/2603.15196}, 
}

@article{esposti_momentum_2024,
	title = {Momentum spectrum of Schwinger pair production in four-dimensional e-dipole fields},
	author = {Degli Esposti, Gianluca and Torgrimsson, Greger},
	journal = {Phys. Rev. D},
	volume = {109},
	issue = {1},
	pages = {016013},
	numpages = {24},
	year = {2024},
	month = {Jan},
	publisher = {American Physical Society},
	doi = {10.1103/PhysRevD.109.016013},
	url = {https://link.aps.org/doi/10.1103/PhysRevD.109.016013}
}

@article{torgrimsson_momentum_2025,
	title = {Momentum correlation in pair production by spacetime dependent fields from scattered wave functions},
	author = {Torgrimsson, Greger},
	journal = {Phys. Rev. D},
	volume = {112},
	issue = {11},
	pages = {116011},
	numpages = {11},
	year = {2025},
	month = {Dec},
	publisher = {American Physical Society},
	doi = {10.1103/j9dq-2hyy},
	url = {https://link.aps.org/doi/10.1103/j9dq-2hyy}
}

@article{torgrimsson_solving_2025,
	title={Solving the Dirac equation on a GPU for strong-field processes in multidimensional background fields},
	author={Torgrimsson, Greger},
	journal={arXiv preprint arXiv:2512.16889},
	year={2025}
}

@article{bras_relative_2026,
	title = {Relative phase and position dependence of assisted electron-positron pair creation in the superposition of strong oscillating electric-field pulses},
	volume = {80},
	issn = {1434-6079},
	url = {https://doi.org/10.1140/epjd/s10053-026-01235-7},
	doi = {10.1140/epjd/s10053-026-01235-7},
	number = {8},
	journal = {The European Physical Journal D},
	author = {Braß, J. and Müller, D. M. and Villalba-Chávez, S. and Krajewska, K. and Müller, C.},
	month = aug,
	year = {2026},
	pages = {126},
}

@article{dirac_quantum_1928,
	title = {The quantum theory of the electron},
	volume = {117},
	url = {https://doi.org/10.1098/rspa.1928.0023},
	doi = {10.1098/rspa.1928.0023},
	number = {778},
	journal = {Proc. A 1},
	publisher = {The Royal Society London},
	author = {Dirac, Paul Adrien Maurice},
	year = {1928},
	pages = {610--624},
}

@article{jangir_carrier-envelope_2026,
	title = {Carrier-envelope phase and pulse-shape effects on vacuum pair production in asymmetric electric fields with bell-shaped envelopes},
	volume = {113},
	url = {https://link.aps.org/doi/10.1103/44k5-hmqt},
	doi = {10.1103/44k5-hmqt},
	number = {9},
	journal = {Phys. Rev. D},
	publisher = {American Physical Society},
	author = {Jangir, Abhinav and Ahmed, Anees},
	month = may,
	year = {2026},
	pages = {096002},
}

@article{fey_momentum_2012,
	title = {Momentum dependence in the dynamically assisted {Sauter}-{Schwinger} effect},
	volume = {85},
	url = {https://link.aps.org/doi/10.1103/PhysRevD.85.025004},
	doi = {10.1103/PhysRevD.85.025004},
	number = {2},
	journal = {Phys. Rev. D},
	publisher = {American Physical Society},
	author = {Fey, Christian and Schützhold, Ralf},
	month = jan,
	year = {2012},
	pages = {025004},
}

@article{aleksandrov_dynamically_2018,
	title = {Dynamically assisted {Schwinger} effect beyond the spatially-uniform-field approximation},
	volume = {97},
	url = {https://link.aps.org/doi/10.1103/PhysRevD.97.116001},
	doi = {10.1103/PhysRevD.97.116001},
	number = {11},
	journal = {Phys. Rev. D},
	publisher = {American Physical Society},
	author = {Aleksandrov, I. A. and Plunien, G. and Shabaev, V. M.},
	month = jun,
	year = {2018},
	pages = {116001},
}

@article{sauter_behavior_1931,
	title = {On the behavior of an electron in a homogeneous electric field in {Dirac}’s relativistic theory},
	volume = {69},
	url = {https://link.springer.com/article/10.1007/BF01339461},
	journal = {Zeit. f. Phys},
	author = {Sauter, Fritz},
	year = {1931},
	pages = {742},
}

@article{blinne_comparison_2016,
	title = {Comparison of semiclassical and {Wigner} function methods in pair production in rotating fields},
	volume = {93},
	url = {https://link.aps.org/doi/10.1103/PhysRevD.93.025014},
	doi = {10.1103/PhysRevD.93.025014},
	number = {2},
	journal = {Phys. Rev. D},
	publisher = {American Physical Society},
	author = {Blinne, Alexander and Strobel, Eckhard},
	month = jan,
	year = {2016},
	pages = {025014},
}

@article{titov_enhanced_2012,
	title = {Enhanced {Subthreshold} e{\textasciicircum}{\textbackslash}mathbf+e{\textasciicircum}{\textbackslash}mathbf{\textbackslash}ensuremath- {Production} in {Short} {Laser} {Pulses}},
	volume = {108},
	url = {https://link.aps.org/doi/10.1103/PhysRevLett.108.240406},
	doi = {10.1103/PhysRevLett.108.240406},
	number = {24},
	journal = {Phys. Rev. Lett.},
	publisher = {American Physical Society},
	author = {Titov, A. I. and Takabe, H. and Kämpfer, B. and Hosaka, A.},
	month = jun,
	year = {2012},
	pages = {240406},
}

@article{bulanov_multiple_2010,
	title = {Multiple {Colliding} {Electromagnetic} {Pulses}: {A} {Way} to {Lower} the {Threshold} of e{\textasciicircum}+e{\textasciicircum}{\textbackslash}ensuremath- {Pair} {Production} from {Vacuum}},
	volume = {104},
	url = {https://link.aps.org/doi/10.1103/PhysRevLett.104.220404},
	doi = {10.1103/PhysRevLett.104.220404},
	number = {22},
	journal = {Phys. Rev. Lett.},
	publisher = {American Physical Society},
	author = {Bulanov, S. S. and Mur, V. D. and Narozhny, N. B. and Nees, J. and Popov, V. S.},
	month = jun,
	year = {2010},
	pages = {220404},
}

@article{di_piazza_barrier_2009,
	title = {Barrier {Control} in {Tunneling} e{\textasciicircum}{\textbackslash}mathbf+{\textbackslash}mathbf{\textbackslash}text{\textbackslash}ensuremath-e{\textasciicircum}{\textbackslash}mathbf{\textbackslash}ensuremath- {Photoproduction}},
	volume = {103},
	url = {https://link.aps.org/doi/10.1103/PhysRevLett.103.170403},
	doi = {10.1103/PhysRevLett.103.170403},
	number = {17},
	journal = {Phys. Rev. Lett.},
	publisher = {American Physical Society},
	author = {Di Piazza, A. and Lötstedt, E. and Milstein, A. I. and Keitel, C. H.},
	month = oct,
	year = {2009},
	pages = {170403},
}

@article{dunne_catalysis_2009,
	title = {Catalysis of {Schwinger} vacuum pair production},
	volume = {80},
	url = {https://link.aps.org/doi/10.1103/PhysRevD.80.111301},
	doi = {10.1103/PhysRevD.80.111301},
	number = {11},
	journal = {Phys. Rev. D},
	publisher = {American Physical Society},
	author = {Dunne, Gerald V. and Gies, Holger and Schützhold, Ralf},
	month = dec,
	year = {2009},
	pages = {111301(R)},
}

@article{heisenberg_folgerungen_1936,
	title = {Folgerungen aus der diracschen theorie des positrons},
	volume = {98},
	url = {https://link.springer.com/article/10.1007/bf01343663},
	number = {11},
	journal = {Zeitschrift für Physik},
	publisher = {Springer},
	author = {Heisenberg, Werner and Euler, Heinrich},
	year = {1936},
	pages = {714--732},
}

@article{schwinger_gauge_1951,
	title = {On {Gauge} {Invariance} and {Vacuum} {Polarization}},
	volume = {82},
	url = {https://link.aps.org/doi/10.1103/PhysRev.82.664},
	doi = {10.1103/PhysRev.82.664},
	number = {5},
	journal = {Phys. Rev.},
	publisher = {American Physical Society},
	author = {Schwinger, Julian},
	month = jun,
	year = {1951},
	pages = {664--679},
}

@article{gies_pair_2005,
	title = {Pair production in inhomogeneous fields},
	volume = {72},
	url = {https://link.aps.org/doi/10.1103/PhysRevD.72.065001},
	doi = {10.1103/PhysRevD.72.065001},
	number = {6},
	journal = {Phys. Rev. D},
	publisher = {American Physical Society},
	author = {Gies, Holger and Klingmüller, Klaus},
	month = sep,
	year = {2005},
	pages = {065001},
}

@article{dunne_worldline_2006,
	title = {Worldline instantons and the fluctuation prefactor},
	volume = {73},
	url = {https://link.aps.org/doi/10.1103/PhysRevD.73.065028},
	doi = {10.1103/PhysRevD.73.065028},
	number = {6},
	journal = {Phys. Rev. D},
	publisher = {American Physical Society},
	author = {Dunne, Gerald V. and Wang, Qing-hai and Gies, Holger and Schubert, Christian},
	month = mar,
	year = {2006},
	pages = {065028},
}

@article{marklund_quantum_2009,
	title = {Quantum vacuum experiments using high intensity lasers},
	volume = {55},
	url = {https://link.springer.com/article/10.1140/epjd/e2009-00169-6#citeas},
	number = {2},
	journal = {The European Physical Journal D},
	publisher = {Springer},
	author = {Marklund, Mattias and Lundin, Joakim},
	year = {2009},
	pages = {319--326},
}

@article{pike_photonphoton_2014,
	title = {A photon–photon collider in a vacuum hohlraum},
	volume = {8},
	url = {https://www.nature.com/articles/nphoton.2014.95},
	number = {6},
	journal = {Nature Photonics},
	publisher = {Nature Publishing Group UK London},
	author = {Pike, OJ and Mackenroth, F and Hill, EG and Rose, SJ},
	year = {2014},
	pages = {434--436},
}

@article{yanovsky_ultra-high_2008,
	title = {Ultra-high intensity-300-{TW} laser at 0.1 {Hz} repetition rate.},
	volume = {16},
	url = {https://opg.optica.org/oe/fulltext.cfm?uri=OE-16-3-2109},
	number = {3},
	journal = {Optics Express},
	publisher = {Optical Society of America},
	author = {Yanovsky, V and Chvykov, V and Kalinchenko, G and Rousseau, P and Planchon, T and Matsuoka, T and Maksimchuk, A and Nees, J and Cheriaux, G and Mourou, G and {others}},
	year = {2008},
	pages = {2109--2114},
}

@article{ringwald_pair_2001,
	title = {Pair production from vacuum at the focus of an {X}-ray free electron laser},
	volume = {510},
	url = {https://doi.org/10.1016/S0370-2693(01)00496-8},
	number = {1-4},
	journal = {Phys. Lett. B},
	publisher = {Elsevier},
	author = {Ringwald, Andreas},
	year = {2001},
	pages = {107--116},
}

@article{popov_pair_1972,
	title = {Pair production in a variable external field (quasiclassical approximation)},
	volume = {34},
	journal = {Soviet Journal of Experimental and Theoretical Physics},
	author = {Popov, VS},
	year = {1972},
	pages = {709},
}

@article{oertel_wkb_2019,
	title = {{WKB} approach to pair creation in spacetime-dependent fields: {The} case of a spacetime-dependent mass},
	volume = {99},
	url = {https://link.aps.org/doi/10.1103/PhysRevD.99.125014},
	doi = {10.1103/PhysRevD.99.125014},
	number = {12},
	journal = {Phys. Rev. D},
	publisher = {American Physical Society},
	author = {Oertel, Johannes and Schützhold, Ralf},
	month = jun,
	year = {2019},
	pages = {125014},
}

@article{linder_pulse_2015,
	title = {Pulse shape dependence in the dynamically assisted {Sauter}-{Schwinger} effect},
	volume = {92},
	url = {https://link.aps.org/doi/10.1103/PhysRevD.92.085009},
	doi = {10.1103/PhysRevD.92.085009},
	number = {8},
	journal = {Phys. Rev. D},
	publisher = {American Physical Society},
	author = {Linder, Malte F. and Schneider, Christian and Sicking, Joachim and Szpak, Nikodem and Schützhold, Ralf},
	month = oct,
	year = {2015},
	pages = {085009},
}

@article{aleksandrov_pair_2020,
	title = {Pair production in temporally and spatially oscillating fields},
	volume = {101},
	url = {https://link.aps.org/doi/10.1103/PhysRevD.101.096009},
	doi = {10.1103/PhysRevD.101.096009},
	number = {9},
	journal = {Phys. Rev. D},
	publisher = {American Physical Society},
	author = {Aleksandrov, Ivan A. and Kohlfürst, Christian},
	month = may,
	year = {2020},
	pages = {096009},
}

@article{aleksandrov_momentum_2017,
	title = {Momentum distribution of particles created in space-time-dependent colliding laser pulses},
	volume = {96},
	url = {https://link.aps.org/doi/10.1103/PhysRevD.96.076006},
	doi = {10.1103/PhysRevD.96.076006},
	number = {7},
	journal = {Phys. Rev. D},
	publisher = {American Physical Society},
	author = {Aleksandrov, I. A. and Plunien, G. and Shabaev, V. M.},
	month = oct,
	year = {2017},
	pages = {076006},
}

@article{lv_role_2018,
	title = {Role of the spatial inhomogeneity on the laser-induced vacuum decay},
	volume = {97},
	url = {https://link.aps.org/doi/10.1103/PhysRevA.97.022515},
	doi = {10.1103/PhysRevA.97.022515},
	number = {2},
	journal = {Phys. Rev. A},
	publisher = {American Physical Society},
	author = {Lv, Q. Z. and {others}},
	month = feb,
	year = {2018},
	pages = {022515},
}

@article{dumlu_stokes_2010,
	title = {Stokes {Phenomenon} and {Schwinger} {Vacuum} {Pair} {Production} in {Time}-{Dependent} {Laser} {Pulses}},
	volume = {104},
	url = {https://link.aps.org/doi/10.1103/PhysRevLett.104.250402},
	doi = {10.1103/PhysRevLett.104.250402},
	number = {25},
	journal = {Phys. Rev. Lett.},
	publisher = {American Physical Society},
	author = {Dumlu, Cesim K. and Dunne, Gerald V.},
	month = jun,
	year = {2010},
	pages = {250402},
}

@article{dunne_worldline_2005,
	title = {Worldline instantons and pair production in inhomogenous fields},
	volume = {72},
	url = {https://doi.org/10.1103/PhysRevD.72.105004},
	number = {10},
	journal = {Phys. Rev. D},
	publisher = {APS},
	author = {Dunne, Gerald V and Schubert, Christian},
	year = {2005},
	pages = {105004},
}

@article{hebenstreit_schwinger_2010,
	title = {Schwinger pair production in space- and time-dependent electric fields: {Relating} the {Wigner} formalism to quantum kinetic theory},
	volume = {82},
	url = {https://link.aps.org/doi/10.1103/PhysRevD.82.105026},
	doi = {10.1103/PhysRevD.82.105026},
	number = {10},
	journal = {Phys. Rev. D},
	publisher = {American Physical Society},
	author = {Hebenstreit, F. and Alkofer, R. and Gies, H.},
	month = nov,
	year = {2010},
	pages = {105026},
}

@phdthesis{hebenstreit_schwinger_2011,
	type = {{PhD} {Thesis}},
	title = {Schwinger effect in inhomogeneous electric fields},
	url = {https://arxiv.org/abs/1106.5965},
	school = {University of Graz},
	author = {Hebenstreit, F.},
	year = {2011},
}

@phdthesis{kohlfurst_electron-positron_2015,
	type = {{PhD} {Thesis}},
	title = {Electron-{Positron} {Pair} {Production} in {Inhomogeneous} {Electromagnetic} {Fields}},
	url = {https://arxiv.org/abs/1512.06082},
	school = {University of Graz},
	author = {Kohlfürst, Christian},
	year = {2015},
}

@article{olugh_pair_2019,
	title = {Pair production in differently polarized electric fields with frequency chirps},
	volume = {99},
	url = {https://doi.org/10.1103/PhysRevD.99.036003},
	number = {3},
	journal = {Phys. Rev. D},
	publisher = {APS},
	author = {Olugh, Obulkasim and Li, Zi-Liang and Xie, Bai-Song and Alkofer, Reinhard},
	year = {2019},
	pages = {036003},
}

@article{chen_asymmetric_2024,
	title = {Asymmetric pulse effects on pair production in chirped electric fields},
	volume = {109},
	url = {https://doi.org/10.1103/PhysRevD.109.076015},
	number = {7},
	journal = {Phys. Rev. D},
	publisher = {APS},
	author = {Chen, Neng-Zhi and Amat, Orkash and Hu, Li-Na and Fan, Hong-Hao and Xie, Bai-Song},
	year = {2024},
	pages = {076015},
}

@article{li_effects_2015,
	title = {Effects of electric field polarizations on pair production},
	volume = {92},
	url = {https://doi.org/10.1103/PhysRevD.92.085001},
	number = {8},
	journal = {Phys. Rev. D},
	publisher = {APS},
	author = {Li, ZL and Lu, D and Xie, BS},
	year = {2015},
	pages = {085001},
}

@article{blinne_pair_2014,
	title = {Pair production in rotating electric fields},
	volume = {89},
	url = {https://link.aps.org/doi/10.1103/PhysRevD.89.085001},
	doi = {10.1103/PhysRevD.89.085001},
	number = {8},
	journal = {Phys. Rev. D},
	publisher = {American Physical Society},
	author = {Blinne, Alexander and Gies, Holger},
	month = mar,
	year = {2014},
	pages = {085001},
}

@article{kluger_quantum_1998,
	title = {Quantum {Vlasov} equation and its {Markov} limit},
	volume = {58},
	url = {https://doi.org/10.1103/PhysRevD.58.125015},
	number = {12},
	journal = {Phys. Rev. D},
	publisher = {APS},
	author = {Kluger, Yuval and Mottola, Emil and Eisenberg, Judah M},
	year = {1998},
	pages = {125015},
}

@article{blaschke_lifting_2015,
	title = {Lifting shell structures in the dynamically assisted {Schwinger} effect in periodic fields},
	volume = {740},
	url = {https://doi.org/10.1016/j.physletb.2014.12.010},
	doi = {10.1016/j.physletb.2014.12.010},
	journal = {Phys. Lett. B},
	author = {Blaschke, D. B. and Dmitriev, V. V. and Röpke, G. and Smolyansky, S. A.},
	year = {2015},
	pages = {335--340},
}

@article{alkofer_pair_2001,
	title = {Pair creation and an {X}-ray free electron laser},
	volume = {87},
	url = {https://doi.org/10.1103/PhysRevLett.87.193902},
	number = {19},
	journal = {Phys. Rev. Lett.},
	publisher = {APS},
	author = {Alkofer, Reinhard and Hecht, MB and Roberts, Craig D and Schmidt, SM and Vinnik, DV},
	year = {2001},
	pages = {193902},
}

@article{hebenstreit_momentum_2009,
	title = {Momentum {Signatures} for {Schwinger} {Pair} {Production} in {Short} {Laser} {Pulses} with a {Subcycle} {Structure}},
	volume = {102},
	url = {https://link.aps.org/doi/10.1103/PhysRevLett.102.150404},
	doi = {10.1103/PhysRevLett.102.150404},
	number = {15},
	journal = {Phys. Rev. Lett.},
	publisher = {American Physical Society},
	author = {Hebenstreit, F. and Alkofer, R. and Dunne, G. V. and Gies, H.},
	month = apr,
	year = {2009},
	pages = {150404},
}

@article{schutzhold_dynamically_2008,
	title = {Dynamically assisted {Schwinger} mechanism},
	volume = {101},
	url = {https://doi.org/10.1103/PhysRevLett.101.130404},
	number = {13},
	journal = {Phys. Rev. Lett.},
	publisher = {APS},
	author = {Schützhold, Ralf and Gies, Holger and Dunne, Gerald},
	year = {2008},
	pages = {130404},
}

@article{nuriman_enhanced_2012,
	title = {Enhanced electron–positron pair creation by dynamically assisted combinational fields},
	volume = {717},
	url = {https://doi.org/10.1016/j.physletb.2012.09.060},
	number = {4-5},
	journal = {Phys. Lett. B},
	publisher = {Elsevier},
	author = {Nuriman, Abdukerim and Xie, Bai-Song and Li, Zi-Liang and Sayipjamal, Dulat},
	year = {2012},
	pages = {465--469},
}

@article{hebenstreit_optimization_2014,
	title = {Optimization of {Schwinger} pair production in colliding laser pulses},
	volume = {739},
	url = {https://doi.org/10.1016/j.physletb.2014.10.056},
	doi = {10.1016/j.physletb.2014.10.056},
	journal = {Phys. Lett. B.},
	author = {Hebenstreit, F. and Fillion-Gourdeau, F.},
	year = {2014},
	pages = {189--195},
}

@article{li_enhanced_2021,
	title = {Enhanced dynamically assisted pair production in spatial inhomogeneous electric fields with the frequency chirping},
	volume = {104},
	url = {https://doi.org/10.1103/PhysRevD.104.036015},
	number = {3},
	journal = {Phys. Rev. D},
	publisher = {APS},
	author = {Li, Lie-Juan and Mohamedsedik, Melike and Xie, Bai-Song},
	year = {2021},
	pages = {036015},
}

@article{blaschke_pair_2006,
	title = {Pair production and optical lasers},
	volume = {96},
	url = {https://doi.org/10.1103/PhysRevLett.96.140402},
	number = {14},
	journal = {Phys. Rev. Lett.},
	publisher = {APS},
	author = {Blaschke, DB and Prozorkevich, AV and Roberts, CD and Schmidt, SM and Smolyansky, SA},
	year = {2006},
	pages = {140402},
}

@article{orthaber_momentum_2011,
	title = {Momentum spectra for dynamically assisted {Schwinger} pair production},
	volume = {698},
	url = {https://doi.org/10.1016/j.physletb.2011.02.053},
	number = {1},
	journal = {Phys. Lett. B},
	publisher = {Elsevier},
	author = {Orthaber, Markus and Hebenstreit, Florian and Alkofer, Reinhard},
	year = {2011},
	pages = {80--85},
}

@article{dumlu_schwinger_2010,
	title = {Schwinger vacuum pair production in chirped laser pulses},
	volume = {82},
	url = {https://doi.org/10.1103/PhysRevD.82.045007},
	number = {4},
	journal = {Phys. Rev. D},
	publisher = {APS},
	author = {Dumlu, Cesim K},
	year = {2010},
	pages = {045007},
}

@article{gong_electron-positron_2020,
	title = {Electron-positron pair production in frequency modulated laser fields},
	volume = {101},
	url = {https://link.aps.org/doi/10.1103/PhysRevD.101.016008},
	doi = {10.1103/PhysRevD.101.016008},
	number = {1},
	journal = {Phys. Rev. D},
	publisher = {American Physical Society},
	author = {Gong, C. and Li, Z. L. and Xie, B. S. and Li, Y. J.},
	month = jan,
	year = {2020},
	pages = {016008},
}

@article{jangir_dynamically_2026,
	title = {Dynamically assisted {Schwinger} pair production in differently polarized electric fields with the frequency chirping},
	volume = {114},
	url = {https://link.aps.org/doi/10.1103/rygd-bkrv},
	doi = {10.1103/rygd-bkrv},
	number = {3},
	journal = {Phys. Rev. D},
	publisher = {American Physical Society},
	author = {Jangir, Abhinav and Ahmed, Anees},
	month = aug,
	year = {2026},
	pages = {036029},
}

@article{olugh_asymmetric_2020,
	title = {Asymmetric pulse effects on pair production in polarized electric fields},
	volume = {8},
	url = {https://doi.org/10.1017/hpl.2020.36},
	journal = {High Power Laser Science and Engineering},
	publisher = {Cambridge University Press},
	author = {Olugh, Obulkasim and Li, Zi-Liang and Xie, Bai-Song},
	year = {2020},
	pages = {e38},
}

@article{keldysh_ionization_1965,
	title = {Ionization in the field of a strong electromagnetic wave},
	volume = {20},
	url = {https://doi.org/10.1142/9789811279461_0008},
	journal = {Sov. Phys. JETP},
	author = {Keldysh, L. V.},
	year = {1965},
	pages = {1307},
}

@article{brezin_pair_1970,
	title = {Pair production in vacuum by an alternating field},
	volume = {2},
	url = {https://doi.org/10.1103/PhysRevD.2.1191},
	number = {7},
	journal = {Phys. Rev. D},
	publisher = {APS},
	author = {Brézin, Edouard and Itzykson, Claude},
	year = {1970},
	pages = {1191},
}

@article{braun_numerical_1999,
	title = {Numerical approach to solve the time-dependent {Dirac} equation},
	volume = {59},
	url = {https://link.aps.org/doi/10.1103/PhysRevA.59.604},
	doi = {10.1103/PhysRevA.59.604},
	number = {1},
	journal = {Phys. Rev. A},
	publisher = {American Physical Society},
	author = {Braun, J. W. and Su, Q. and Grobe, R.},
	month = jan,
	year = {1999},
	pages = {604--612},
}

@article{li_study_2021,
	title = {Study of pair production in inhomogeneous two-color electric fields using the computational quantum field theory},
	volume = {103},
	url = {https://link.aps.org/doi/10.1103/PhysRevD.103.116018},
	doi = {10.1103/PhysRevD.103.116018},
	number = {11},
	journal = {Phys. Rev. D},
	publisher = {American Physical Society},
	author = {Li, Z. L. and Gong, C. and Li, Y. J.},
	month = jun,
	year = {2021},
	pages = {116018},
}

@article{abdukerim_effects_2013,
	title = {Effects of laser pulse shape and carrier envelope phase on pair production},
	volume = {726},
	url = {https://doi.org/10.1016/j.physletb.2013.09.014},
	number = {4-5},
	journal = {Phys. Lett. B},
	publisher = {Elsevier},
	author = {Abdukerim, Nuriman and Li, Zi-Liang and Xie, Bai-Song},
	year = {2013},
	pages = {820--826},
}

@article{dumlu_interference_2011,
	title = {Interference effects in {Schwinger} vacuum pair production for time-dependent laser pulses},
	volume = {83},
	url = {https://doi.org/10.1103/PhysRevD.83.065028},
	number = {6},
	journal = {Phys. Rev. D},
	publisher = {APS},
	author = {Dumlu, Cesim K and Dunne, Gerald V},
	year = {2011},
	pages = {065028},
}

@article{dumlu_complex_2011,
	title = {Complex worldline instantons and quantum interference in vacuum pair production},
	volume = {84},
	url = {https://doi.org/10.1103/PhysRevD.84.125023},
	number = {12},
	journal = {Phys. Rev. D},
	publisher = {APS},
	author = {Dumlu, Cesim K and Dunne, Gerald V},
	year = {2011},
	pages = {125023},
}

@article{hebenstreit_particle_2011,
	title = {Particle {Self}-{Bunching} in the {Schwinger} {Effect} in {Spacetime}-{Dependent} {Electric} {Fields}},
	volume = {107},
	url = {https://link.aps.org/doi/10.1103/PhysRevLett.107.180403},
	doi = {10.1103/PhysRevLett.107.180403},
	number = {18},
	journal = {Phys. Rev. Lett.},
	publisher = {American Physical Society},
	author = {Hebenstreit, F. and Alkofer, R. and Gies, H.},
	month = oct,
	year = {2011},
	pages = {180403},
}

@article{ababekri_effects_2019,
	title = {Effects of finite spatial extent on {Schwinger} pair production},
	volume = {100},
	url = {https://link.aps.org/doi/10.1103/PhysRevD.100.016003},
	doi = {10.1103/PhysRevD.100.016003},
	number = {1},
	journal = {Phys. Rev. D},
	publisher = {American Physical Society},
	author = {Ababekri, Mamutjan and Xie, Bai-Song and Zhang, Jun},
	month = jul,
	year = {2019},
	pages = {016003},
}

@article{bake_vacuum_2025,
	title = {Vacuum pair production in spatially asymmetric time-oscillating electric fields},
	volume = {112},
	url = {https://link.aps.org/doi/10.1103/myr3-ql9h},
	doi = {10.1103/myr3-ql9h},
	number = {1},
	journal = {Phys. Rev. D},
	publisher = {American Physical Society},
	author = {Bake, Mamat Ali and Olugh, Obulkasim},
	month = jul,
	year = {2025},
	pages = {016030},
}

@article{ababekri_chirp_2020,
	title = {Chirp effects on pair production in oscillating electric fields with spatial inhomogeneity},
	volume = {810},
	url = {https://www.sciencedirect.com/science/article/pii/S0370269320306183},
	journal = {Physics Letters B},
	publisher = {Elsevier},
	author = {Ababekri, Mamutjan and Dulat, Sayipjamal and Xie, BS and Zhang, Jun},
	year = {2020},
	pages = {135815},
}

@article{osman_efficient_2023,
	title = {Efficient vacuum pair production in frequency chirped electric fields with spatiotemporal inhomogeneity},
	volume = {55},
	url = {https://www.sciencedirect.com/science/article/pii/S2211379723009889},
	journal = {Results in Physics},
	publisher = {Elsevier},
	author = {Osman, Emin and Bai, Jinhui and Chen, Zhaoyuan and Bake, Mamat Ali},
	year = {2023},
	pages = {107195},
}

@article{mohamedsedik_phase_2023,
	title = {Phase effect on and symmetry of pair production in inhomogeneous electric fields with chirping},
	volume = {138},
	number = {4},
	journal = {The European Physical Journal Plus},
	publisher = {Springer},
	author = {Mohamedsedik, Melike and Li, Lie-Juan and Wang, Li and Amat, Orkash and Hu, Li-Na and Xie, BS},
	year = {2023},
	pages = {316},
}

@article{di_piazza_extremely_2012,
	title = {Extremely high-intensity laser interactions with fundamental quantum systems},
	volume = {84},
	url = {https://link.aps.org/doi/10.1103/RevModPhys.84.1177},
	doi = {10.1103/RevModPhys.84.1177},
	number = {3},
	journal = {Rev. Mod. Phys.},
	publisher = {American Physical Society},
	author = {Di Piazza, A. and Müller, C. and Hatsagortsyan, K. Z. and Keitel, C. H.},
	month = aug,
	year = {2012},
	pages = {1177--1228},
}

@article{fedotov_advances_2023,
	title = {Advances in {QED} with intense background fields},
	volume = {1010},
	url = {https://www.sciencedirect.com/science/article/pii/S0370157323000352},
	journal = {Physics Reports},
	publisher = {Elsevier},
	author = {Fedotov, A and Ilderton, A and Karbstein, F and King, Ben and Seipt, D and Taya, H and Torgrimsson, Greger},
	year = {2023},
	pages = {1--138},
}

@misc{eli,
	title = {https://www.eli-laser.eu/},
	url = {https://www.eli-laser.eu/},
}

@misc{xfel,
	title = {https://www.xfel.eu/},
	url = {https://www.xfel.eu/},
}

@incollection{dunne_heisenberg-euler_2005,
	title = {{HEISENBERG}-{EULER} {EFFECTIVE} {LAGRANGIANS}: {BASICS} {AND} {EXTENSIONS}},
	url = {https://www.worldscientific.com/doi/abs/10.1142/9789812775344_0014},
	doi = {10.1142/9789812775344_0014},
	booktitle = {From {Fields} to {Strings}: {Circumnavigating} {Theoretical} {Physics}},
	author = {DUNNE, GERALD V.},
	year = {2005},
	pages = {445--522},
}

@article{li_nonperturbative_2015,
	title = {Nonperturbative signatures in pair production for general elliptic polarization fields},
	volume = {110},
	url = {https://iopscience.iop.org/article/10.1209/0295-5075/110/51001/meta},
	number = {5},
	journal = {Europhysics Letters},
	publisher = {EDP Sciences, IOP Publishing and Società Italiana di Fisica},
	author = {Li, ZL and Lu, D and Xie, BS and Shen, BF and Fu, LB and Liu, J},
	year = {2015},
	pages = {51001},
}

@article{olugh_frequency_2026,
	title = {Frequency chirp-induced enhancement of electron–positron pair production in polarized laser fields},
	volume = {35},
	url = {https://doi.org/10.1088/1674-1056/ae0a3c},
	doi = {10.1088/1674-1056/ae0a3c},
	number = {4},
	journal = {Chinese Physics B},
	publisher = {Chinese Physical Society and IOP Publishing Ltd},
	author = {Olugh, Obulkasim},
	month = apr,
	year = {2026},
	pages = {045202},
}
%%%%%%%%%%%%%%%%%%%%%%%%%%%%%%%%%%%%%%%%%%%%%%%%%%%%%%%%%%%%%%%%%
\end{document}